\documentclass{ws-rv975x65}
\usepackage{subfigure}     
\usepackage{ws-rv-thm}     
\usepackage{ws-rv-van}     
\usepackage{aas_macros}
\usepackage{amsmath}
\usepackage{xcolor}
\usepackage[colorlinks=true]{hyperref}
\usepackage{siunitx}
\makeindex

\newcommand{\msg}[1]{\textcolor{blue}{#1}}

\newcommand{\ie}{{\it i.e.}}
\newcommand{\eg}{{\it e.g.}}

\crop[off]
\begin{document}

\chapter[Astrometric Interferometry]{Astrometric Interferometry}\label{ra_ch1}

\author{Michael J. Ireland}
\address{Research School of Astronomy and Astrophysics,\\
Australian National University, Canberra, ACT 2611, Australia, \\
michael.ireland@anu.edu.au}

\author[M. Ireland and J. Woillez]{Julien Woillez}
\address{European Southern Observatory,\\
Karl-Schwarzschild-Str. 2, Garching bei M\"unchen, Germany,\\
jwoillez@eso.org}

\begin{abstract}
Astrometry is a powerful technique in astrophysics to measure three-dimensional positions of stars
and other astrophysical objects, including exoplanets and the gravitational influence they have on each other. 
Interferometric astrometry is presented here as just one in a suite of powerful astrometric techniques, which
include space-based, seeing-limited and wide-angle adaptive optics techniques. Fundamental limits are 
discussed, demonstrating that even ground-based techniques have the capability for astrometry at the 
single micro-arcsecond level, should sufficiently sophisticated instrumentation be constructed for both
the current generation of single telescopes and long-baseline optical interferometers.
\end{abstract}
\body

\section{The Promise of Astrometry}
\label{Sec:Introduction}

Accurate measurements of the time-dependent angular positions of stars and other astrophysical objects are among the most fundamental measurements in astrophysics. Both wide-angle and narrow-angle astrometric measurements have a long history of advancing a broad range of astrophysical topics \cite{Perryman2012}. Tycho Brahe's sextant produced the first accurate measurements of the planets, enabling Kepler's laws and leading to Newtonian mechanics and gravitation. These measurements were {\em wide-angle} measurements, where an instrument has to slew to different stars, making measurements at independent times. The famous measurement of the gravitational deflection of light \cite{Dyson+1920} was an example of a {\em narrow-angle} measurement, where a target and many reference stars are observed simultaneously, cancelling out some instrumental effects. More recently, seeing-limited narrow-angle astrometric measurements have focused on parallaxes of nearby stars and detections of exoplanets through their gravitational influence on their host stars. The semi-amplitude of the astrometric signal of an exoplanet is given by:
\begin{align}
\Delta \theta &\approx \Big( \frac{M_P}{M_S}\Big) \Big( \frac{a_p}{1\,{\rm AU}}\Big) \varpi \, , 
\end{align}
where $M_P$ is the mass of the planet, $M_S$ is the mass of the star, $a_p$ is the semi-major axis of the planet and $\varpi$ is the star's parallax. This means that measuring the astrometric signal from an exoplanet with 1\,AU separation is at least $M_S/M_P$ (about 10$^3$ for $M_P=1$ Jupiter mass) times more difficult than measuring stellar parallaxes.

Many recent ground-based astrometric measurements have been controversial, including the famous example of the debunked planet around Barnard's star \cite{vandeKamp1963}, and more recently the debunked planet around VB~10~\cite{Pravdo&Shaklan2009,Bean+2010}. Other measurements are less controversial, for example infrared parallaxes of field brown dwarfs \cite{Liu+2016}, which provide a determination of the luminosity and fundamental parameters of the coolest objects in the galaxy, not possible by other means.

\begin{figure}
\includegraphics[width=0.55\textwidth]{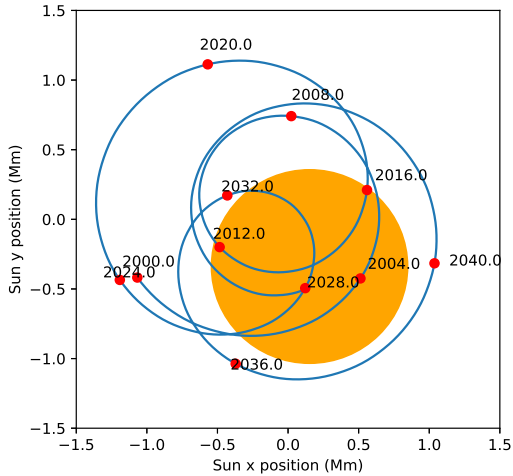}
\caption{Motion of the Sun's barycenter due to perturbations of solar-system planets. The Sun's disk is the orange circle for reference. Scaling between these linear units and angular units for a solar system analog can easily be done, with the solar disk diameter being equivalent to $\sim$1/107th of the stellar parallax.  See electronic edition for a color version of this figure.}
\label{fig:SunMotion}
\end{figure}

Space-based wide-angle astrometric measurements are a cornerstone of modern astrophysics. The HIPPARCOS spacecraft solidified distance scales within the Galaxy, and, via standard candles, the Universe. The Gaia spacecraft is in the process of very significantly extending this work \cite{Perryman+2001}. Depending on the mission length of Gaia and the fraction of a detected exoplanet orbit regarded as a secure detection, Gaia could discover up to 70,000 planets around other stars via astrometry \cite{Perryman+2014}.

The highest precision astrometry, however, requires large apertures or long baselines, and is arguably not best achieved from space. The Gravity instrument at the Very Large Telescope Interferometer (VLTI) \cite{GravityCollaboration+2017} achieves very-narrow-angle astrometric precisions comparable to Gaia, but on much fainter stars. In order to understand the potential role of astrometric interferometry in the broad context of astrometry, we must first understand the fundamental limitations of the technique. With this motivation, this chapter first begins with describing fundamental astrophysical limits to astrometric measurements for given aperture areas and interferometric baselines, and moves on to limitations provided by the Earth's atmosphere in increasing level of detail. Seeing-limited, adaptive-optics assisted and long baseline interferometric techniques are discussed together, with a telescope diameter $D$ or an interferometric baseline $B$ used almost interchangeably. 

It is assumed that the reader of this chapter has a basic understanding of atmospheric turbulence theory (\ie, Fried's coherence length and coherence time),\footnote{See Chapter 1 of this Volume for a brief discussion of atmospheric turbulence theory.} 
narrow-field optical interferometry and adaptive optics. Electric fields are approximated in this chapter as scalar wavefronts, and we consider vectors and angles in a plane perpendicular to the vector between a telescope and the center of the observed field. Finally, the practical difficulties in building real astrometric instruments is briefly discussed in principle, leading to a discussion of past, present and possible future astrometric interferometers.

\section{Astrometric Precision without the Earth's Atmosphere}
\label{Sec:Photon}

For a star of AB magnitude $m_{\lambda}$ observed with a fractional bandwidth $\Delta \lambda/\lambda$, a total aperture area $A$, integration time $\Delta T$ and overall instrument throughput $\eta$, the number of photons detected is given by:
\begin{align}
N_p &= 5.48 \times 10^{10} \eta A \Delta T \frac{\Delta \lambda}{\lambda} 10^{-0.4 m_\lambda}.
\end{align}
%
An AB magnitude is roughly the same as a Vega magnitude at visible wavelengths, and reaches $m_{\rm Vega}-m_{AB}=1$ at \SI{\sim1.28}{\micro\meter}\cite{Tokunaga&Vacca2005}. Armed with the knowledge of how many photons are expected from astrophysical sources, we can use maximum likelihood estimation under the assumption of a perfectly sampled point-spread function to determine a target shot-noise limited astrometric error. For image-plane detection with a point-spread function $f(\alpha, \beta)$ for angular co-ordinates $\alpha$ and $\beta$, the photon-limited centroid error from maximum likelihood estimation can be easily shown to be:
\begin{align}
\sigma_\alpha    &= \frac{\alpha_{\rm psf}}{\sqrt{N_p}}, {\rm~where}  \label{eqnPhotonNoise} \\
\alpha_{\rm psf}^2 &= \frac{1}{\iint  \frac{\partial f}{\partial \alpha}^2 (\alpha, \beta) / f(\alpha, \beta) d\alpha d\beta}.
\end{align}
Here the point spread function $f$ is assumed to be normalized (\ie, an integral of unity), and $N_p$ is the total number of photons recorded in the image. This type of relationship can be derived directly from quantum mechanics \cite{Lindegren2010}, or from differentiating the logarithm of the maximum likelihood for a 
Taylor-expanded point-spread function in the limit of small values of $\alpha$ and small pixels. 
In practice, uncertainties will be close to these limits as long as the point-spread function is critically sampled (more than 2 pixels per FWHM),  the core of the point spread function is significantly brighter than any background, and detector readout noise is negligible compared to any shot noise.

The effective point-spread function size $\alpha_{\rm psf}$ is $\sim 0.33\lambda/D$ for a fully-filled diffraction-limited aperture of diameter $D$, $\sim 0.16 \lambda/B = \lambda/2\pi B$ for a 2-aperture image-plane interferometer of baseline $B$, or $\sim 0.53 \theta_{\rm s}$ for a seeing-limited image with seeing full-width-half-maximum $\theta_{\rm s}$. Note that interferometers do typically have lower throughput than adaptive optics systems, which means that for the same number of photons, an interferometer of baseline $B$ is nearly equivalent to a telescope of diameter $D$. 
\Eref{eqnPhotonNoise} then means that we can approximately consider an interferometer to be superior to a telescope from a single-star photon-limited perspective whenever $D_{\rm int} B_{\rm int} \gtrsim D_{\rm tel}^2$. This means that the VLTI with the Auxiliary Telescopes should have better photon-limited astrometric precision than an individual VLT Unit Telescope, but the E-ELT should have better photon-limited precision than even the VLTI with the Unit Telescopes. 

\begin{figure}
\includegraphics[width=0.49\textwidth]{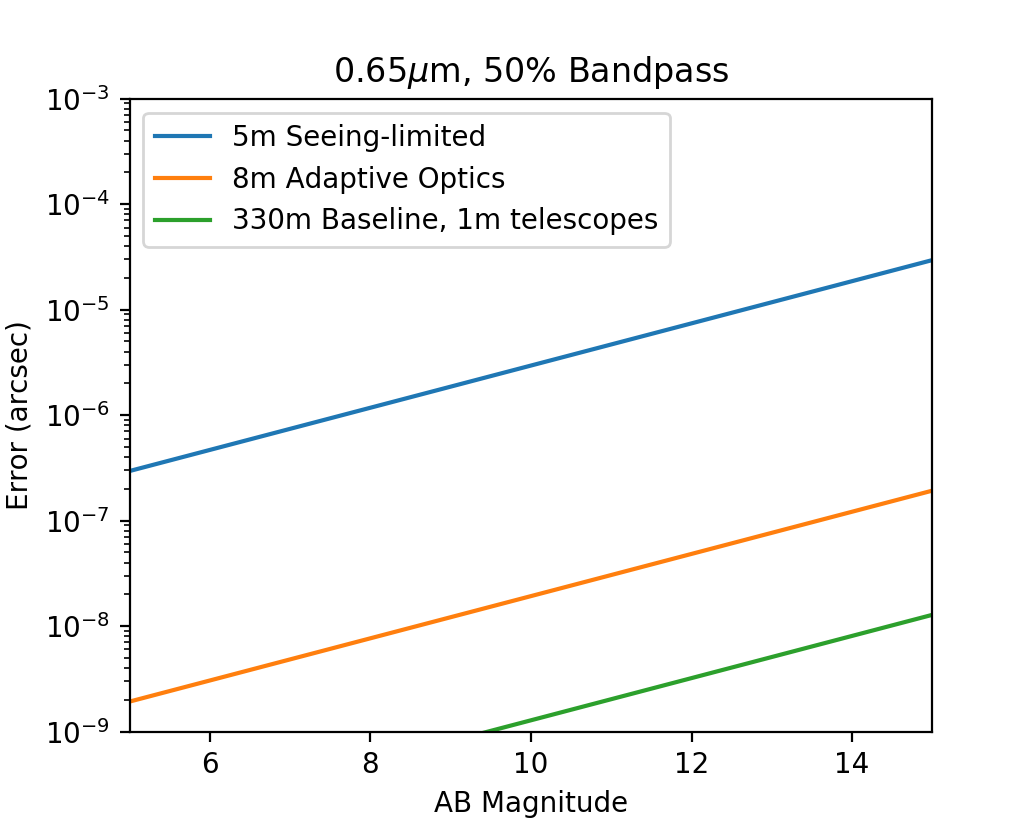} 
\includegraphics[width=0.49\textwidth]{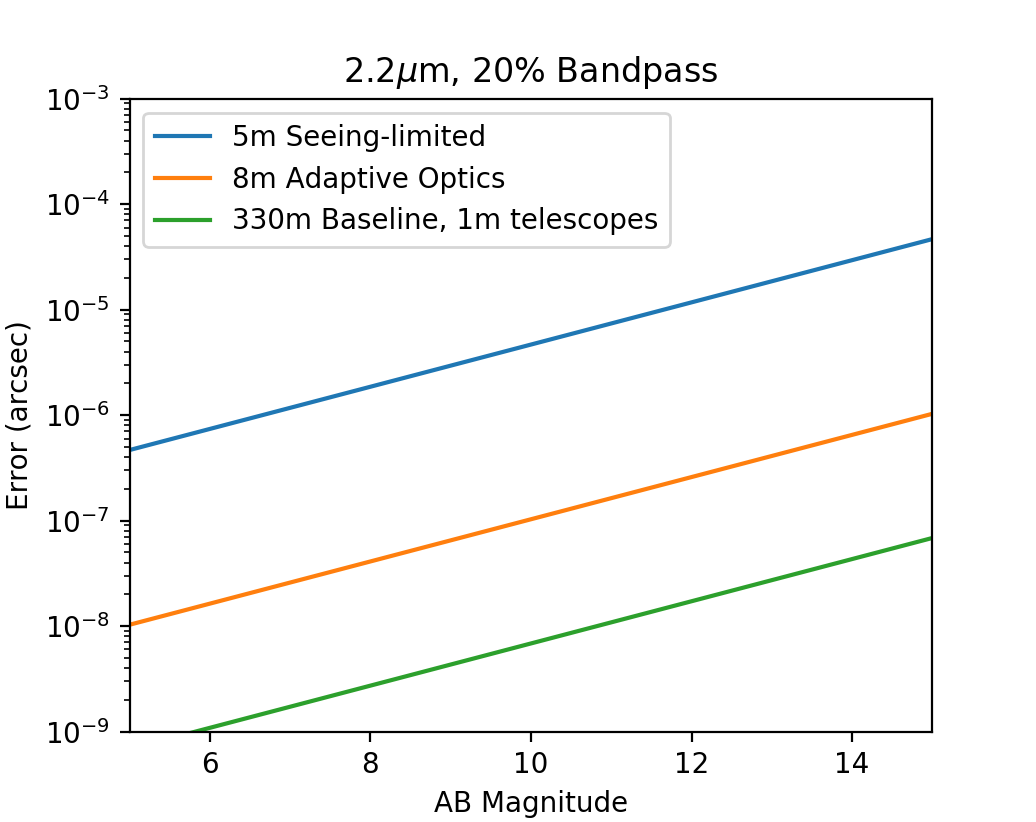} 
\caption{Astrometric uncertainties limited by photon shot noise as a function of AB magnitude, assuming an integration time-throughput product of 10 minutes. For large adaptive optics equipped telescopes or long baseline interferometers, shot noise in centroid measurement is not a dominant error term.  See electronic edition for a color version of this figure.}
\label{figPhotonNoise}
\end{figure}

These fundamental limits correspond to very small angles for typical astrophysical targets, as shown in \fref{figPhotonNoise}. Although incomplete (\eg, not taking into account background noise), this demonstrates that less fundamental effects will limit astrometric precision, such as instrumental imperfections (\eg, imaging system distortions) and high-altitude atmospheric turbulence.

\section{The Effect of the Earth's Atmosphere on Astrometry}
\label{Sec:Atmosphere}


In the literature, there are numerous discussions of atmospheric turbulence profiles and appropriate multi-layer atmosphere models.\footnote{See Chapter 1 of this Volume for a brief discussion.} In order to simplify analysis of a complex atmosphere with many layers, we will consider atmospheric turbulence from a single layer with at an {\em effective height} $\bar{h}$, moving as a frozen flow with an {\em effective velocity} $\bar{v}$. Both of these quantities come from turbulence-weighted averages throughout the atmosphere of each quantity to the 5/3 power:
\begin{align}
\bar{h} &= \Big(\frac{\int C_n^2(h) h^{5/3} dh}{\int C_n^2(h) dh}\Big)^{3/5} \\
\nonumber\\
\bar{v} &= \Big( \frac{\int C_n^2(h) v(h)^{5/3} dh}{\int C_n^2(h) dh} \Big)^{3/5}
\end{align}

Turbulence is stronger towards the lower atmospheric layers, so a typical effective turbulence height $\bar{h}$ at a mountaintop site might be \SI{2.5}{\kilo\meter}, with an effective wind velocity  $\bar{v}$ of \SIrange{15}{20}{\meter\per\second}\cite{Sarazin&Tokovinin2002}. By simplifying the atmospheric profile to a single layer, we can also consider our single layer going at an average speed $\bar{v}$ to have a velocity vector $\bar{\boldsymbol{v}}$. This assumption of course changes some details of this chapter, and any results that require a lucky turbulent layer velocity direction should not be viewed as realistic.

\subsection{Single Star Astrometry}
\label{SSec:SingleStarAstrometry}

In the case of single-star (sometimes called wide-angle) astrometry from the ground, the Earth's atmosphere has two key effects - it increases the photon-limited uncertainty due to a larger point-spread function (\sref{Sec:Photon}), and it shifts the star image to and fro via the {\em tip/tilt} mode of a single aperture, or the {\em piston} model of an interferometer. We can approximate the power spectral density of a tip/tilt mode of a single telescope at low frequencies as, \eg
\cite{tenBrummelaar1996,Glindemann1997,Glindemann2011},
%
\begin{align}
P(f) &= 0.096 (r_0/\bar{v})^{1/3}(\lambda/r_0)^2 f^{-2/3} (1 - e^{-(f L_0 / 2\pi \bar{v})^2})) {\rm~[rad}^2/{\rm Hz]}, {\rm~or} \label{eqnExponentialTipTilt} \\
P(f) &= 0.096 (r_0/\bar{v})^{1/3}(\lambda/r_0)^2 (f^2 + (\bar{v}/L_0)^2)^{-1/3} {\rm~[rad}^2/{\rm Hz]} \, , 
\label{eqnvonKarmannTipTilt}
\end{align}
up to a cut-off frequency $f_c=0.24\bar{v}/D$. Here $r_0$ is the Fried coherence length,\footnote{See Chapter 1 of this Volume.} $\lambda$ is the observing wavelength, $\bar{v}$ is the effective wind velocity and $L_0$ is the turbulence outer scale in an exponential model (\eref{eqnExponentialTipTilt}) or the often used von Karmann model (\eref{eqnvonKarmannTipTilt}). 

The effect of a large aperture is to change the cutoff frequency, reducing the instantaneous rms tilt,  {\em but it does not affect the long-integration time astrometric uncertainty}. We evaluate this uncertainty by integrating this power spectral density multiplied by a sinc$^2$ function, \ie,
\begin{align}
\sigma_\theta &= \sqrt{2 \int_0^{f_c} P(f) {\rm sinc}^2(\pi f \Delta T) df }\\
 &\approx 0.55 (r_0/\bar{v})^{1/6}(\lambda/r_0) \Delta T^{-1/6} {\rm~for~}L_0>\bar{v}\Delta T>D \, .
\end{align}
Inserting typical values of $L_0=\SI{100}{\meter}$, $\bar{v}=\SI{5}{\meter\per\second}$, and $r_0=\SI{0.1}{\meter}$ at $\lambda=\SI{0.5}{\micro\meter}$, this gives a \SI{0.18}{\arcsecond} uncertainty after a \SI{20}{\second} integration. For longer integration times, uncertainties decline rapidly, but are very dependent on the details of the turbulence outer scale. It is quite clear that star motions from bulk atmospheric flows do not cancel out when one star is observed at a time. This is one of the reasons that ground-based single-star astrometry has struggled to overcome atmospheric limitations and produce uncertainties below \SI{0.02}{\arcsecond} \cite{Benson+2010}.

\subsection{Dual Star Astrometry}
\label{SSec:DualStarAstrometry}

\begin{figure}
\centering
\def\svgwidth{0.9\textwidth}
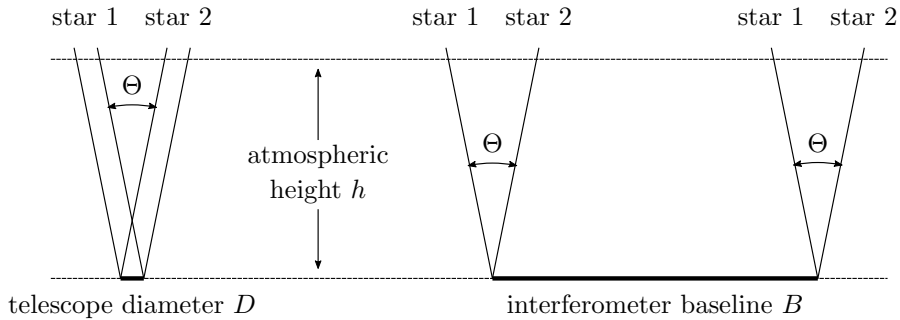
\caption{Dual-star astrometry in the narrow-angle (left, $\Theta h \ll D$) and very-narrow-angle (right, $\Theta h \gg B$) regimes.  After Ref.~\refcite{Shao&Colavita1992}.}
\label{figVNA}
\end{figure}

When at least one additional star can be observed simultaneously with a target, relative astrometric uncertainties are significantly reduced. This is the case for seeing-limited imaging, adaptive optics assisted observations and long-baseline interferometry. As shown in \fref{figVNA}, in dual-star astrometry (often called {\em narrow-angle} astrometry), pupil-plane aberrations in telescopes, including piston phase offsets in a long baseline interferometer, cancel out for both stars observed. In the special case of {\em very-narrow-angle} astrometry\cite{Shao&Colavita1992,Lindegren1980}, aberrations from the upper atmosphere also partially cancel out.

A key atmospheric parameter in determining the uncertainties in dual-star astrometry is the isoplanatic angle, which is the angle at which the RMS phase difference between two thin beams traveling through the atmosphere is 1 radian. This is given by \cite{Glindemann2011}:
\begin{align}
\theta_0 &= 0.314 \frac{r_0}{\bar{h}}.
\end{align}
Ground layer turbulence changes $r_0$ and $\bar{h}$ equally, so does not affect $\theta_0$ and can be ignored in astrometry (assuming that AO systems and fringe trackers still function). For typical \SI{1}{\arcsecond} seeing at visible wavelengths, with turbulence spread throughout the atmosphere at an effective altitude of 2.5\,km, we obtain $\theta_0=\SI{2.6}{\arcsecond}$ in the visible, and \SI{16}{\arcsecond} in the K filter (\SI{2.2}{\micro\meter}).  Isoplanatic angles are often used as a proxy for how far off axis guide stars can be to act as a coherent phase reference, for phase referenced astrometry or off-axis adaptive optics. 

As apertures in both single-telescope astrometry and interferometric astrometry are typically larger than the Fried coherence length $r_0$, observations can average over larger patches of atmosphere, filtering out high spatial frequency components of the atmosphere and obtaining larger coherent fields of view.  This results in more advanced terminology, such as the {\em isokinetic angle} in laser-guide star adaptive optics, and the {\em isopistonic angle} in long baseline interferometry. We will ignore these effects here, and simply assume that the tilt or piston is measured for two stars separated by some angle. This is effectively an assumption that either the interferometric target flux is bright enough for simultaneous fringe tracking on two objects, or the aperture size is large enough so that off-axis phase referencing is viable.  In addition, it is an assumption that an adaptive optics system can provide sufficiently high off-axis Strehl ratio in order to meet either of these conditions. These assumptions also limit the validity of our discussion to the long exposure time regime; fluctuations on short exposures will depend on aperture size, fringe tracking and adaptive optics details, but will average out. Finally, we will also use terminology of baseline rather than telescope diameter in discussing the effects of phase fluctuations on astrometric uncertainties, but note that telescope diameter can be roughly substituted for baseline in the case of single aperture measurements.

\begin{figure}
\centering
\includegraphics[width=0.7\textwidth]{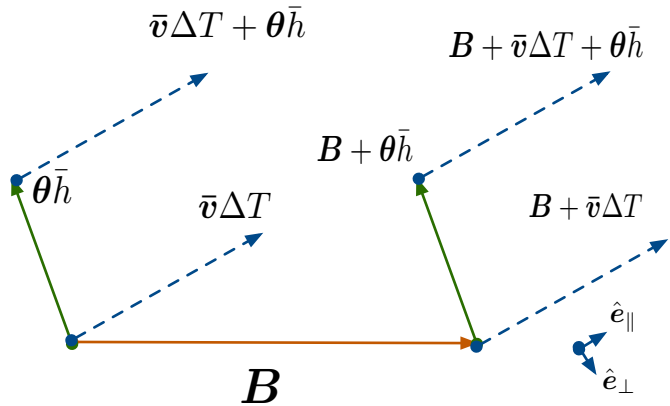}
\caption{Illustration of vectors $\boldsymbol{\theta}\bar{h}$, $\mathbf{B}$ and $\bar{\boldsymbol{v}} \Delta T$ on the wavefront $\varphi(\boldsymbol{x})$ at an effective turbulence height $\bar{h}$. The Taylor hypothesis enables the spatio-temporal correlation to be approximated as a spatial correlation between the wavefront at 8 vector positions. Unit vectors parallel 
($\hat{\boldsymbol{e}}_\parallel$) and perpendicular ($\hat{\boldsymbol{e}}_\perp$) to the wind direction are also shown.}
\label{figVectors}
\end{figure}

With these assumptions, we can reduce the question of astrometric uncertainties to spatio-temporal correlations on a wavefront at height $\bar{h}$. The measurement of differential piston between two stars is given by:
\begin{align}
\frac{\Delta \theta \lambda}{2\pi} &= |B| \Big( [ \varphi(\mathbf{B}) - \varphi(0) ] - [ \varphi(\mathbf{B} + \boldsymbol{\theta}\bar{h}) - \varphi(\boldsymbol{\theta}\bar{h}) ] \Big).
\label{eqnPhaseDifference}
\end{align}
Here $\Delta \theta$ is the instantaneous astrometric error corresponding to a phase uncertainty on baseline $\mathbf{B}$. We consider fluctuations in $\Delta \theta$ as the vectors $\mathbf{B}$ and $\boldsymbol{\theta}$ move around the wavefront by the spatial co-ordinate $\bar{\boldsymbol{v}}\Delta T$, or equivalently its reciprocal space vector $\boldsymbol{\kappa}$.

It should be instantly clear that this uncertainty is zero for the lowest-order wavefront spatial frequencies, which correspond to a tilt on the wavefront $\varphi$. High spatial frequencies, with $|\boldsymbol{\kappa}|>1/|\theta|\bar{h}$ and $|\boldsymbol{\kappa}|>1/|B|$ will have twice the variance of a fringe on a single baseline, and will average out relatively quickly, while intermediate spatial frequencies are partially cancelled out in the dual star mode.

We can consider the power spectrum of $\Delta \theta$ by recognizing that the sums and differences of wavefront positions expressed in \eref{eqnPhaseDifference} are equivalent to convolution in the Fourier domain. We then arrive at the two-dimensional spatial power spectrum of differential astrometry fluctuations \cite{Shao&Colavita1992,Glindemann2011}:
\begin{align}
\Theta (\boldsymbol{\kappa}) &= 0.00928 |B|^{-2} \lambda^2 r_0^{-5/3} |\boldsymbol{\kappa}|^{-11/3} \sin^2(\pi\mathbf{B} \cdot \boldsymbol{\kappa}) \sin^2(\pi \bar{h}\boldsymbol{\theta} \cdot \boldsymbol{\kappa}).
\end{align}
%
Note that the term $\lambda^2 r_0^{-5/3}$ can also be written as $0.0083 \theta_{\rm seeing, \SI{0.5}{\micro\meter}}^{5/3}$, where $\theta_{\rm seeing, \SI{0.5}{\micro\meter}}$ is the visible seeing disk full-width-half-maximum, and so is independent of wavelength.

Converting this spatial 
spectrum
to a temporal spectrum involves considering the evolution of this differential astrometric signal as the wavefront moves past along the vectors $\bar{\boldsymbol{v}}\Delta{T}$ after a time $\Delta T$ (see \fref{figVectors}). This is in turn an integral in  
$\boldsymbol{\kappa}$-space \cite{Conan+1995}, along the dimensionless unit-vector direction $\hat{\boldsymbol{e}}_\perp$, which is perpendicular to $\bar{\boldsymbol{v}}$, enabled by the decomposition $\boldsymbol{\kappa} = (f/v) \hat{\boldsymbol{e}}_\parallel + \kappa_\perp \hat{\boldsymbol{e}}_\perp$:
\begin{align}
\Theta (f) &= \frac{1}{\bar{v}} \int \Theta \Big( \frac{f}{\bar{v}}\hat{\boldsymbol{e}}_\parallel + \kappa_\perp \hat{\boldsymbol{e}}_\perp\Big) d\kappa_\perp.
\end{align}

\begin{figure}
\centering
\includegraphics[width=0.9\textwidth]{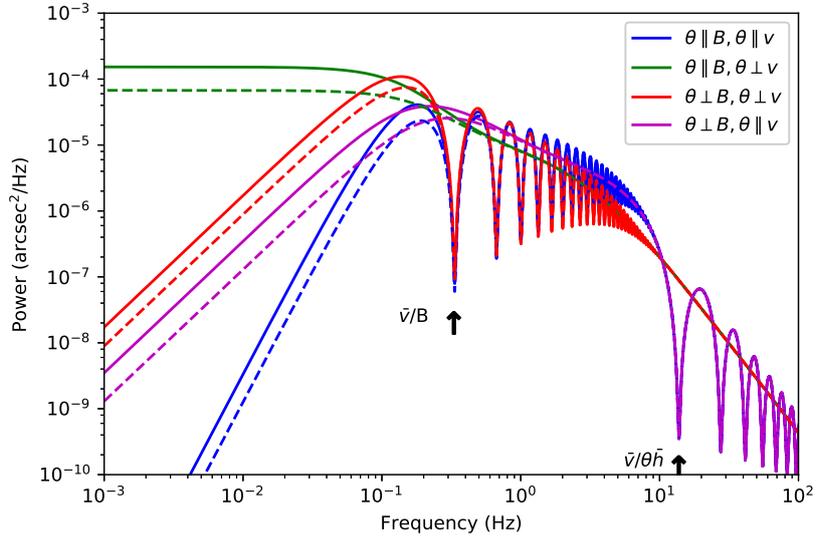}
\caption{The temporal power spectrum of differential astrometry, for point-like apertures and the four limiting geometrical cases, with the angle between the two stars $\boldsymbol{\theta}$ being either parallel or perpendicular to the baseline and wind direction. Parameters for the calculation are: \SI{30}{\meter} baseline $B$, \SI{1}{\arcminute} star separation, \SI{10}{\meter\per\second} wind speed, \SI{2.5}{\kilo\meter} effective turbulence height, \SI{2.2}{\micro\meter} wavelength and \SI{0.6}{\meter} Fried coherence length $r_0$ at this wavelength. Dashed lines show the same calculation with a von Karmann outer scale length of $L_0=\SI{60}{\meter}$. Key knee/null frequencies are shown, and in all cases the total RMS differential angular error is 6 -- 7 milli-arcseconds (\SI{2.7}{\radian} of fringe phase).  See electronic edition for a color version of this figure.}
\label{figDualStarPS}
\end{figure}

\begin{figure}
\centering
\includegraphics[width=0.9\textwidth]{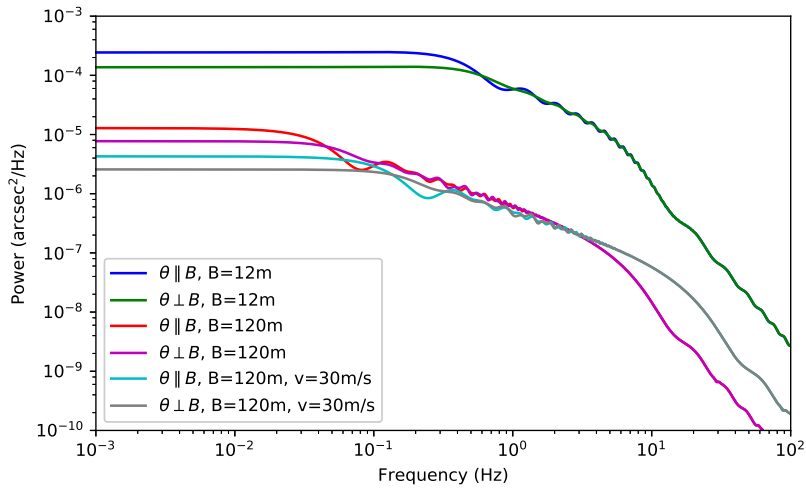}
\caption{The same as \fref{figDualStarPS}, except averaged over all wind directions, for \SI{12}{\meter} and \SI{120}{\meter} baseline lengths and including a higher wind velocity as well as the default of \SI{10}{\meter\per\second}. See electronic edition for a color version of this figure.}
\label{figDualStarRotAvgPS}
\end{figure}

This integral can be computed numerically, and has been calculated for the four limiting geometries covering the relative directions of $\mathbf{B}$, $\boldsymbol{\kappa}$ and $\boldsymbol{\theta}$, as shown in \fref{figDualStarPS}.  In all cases, the RMS astrometric uncertainty is \SI{\sim6} milli-arcseconds, although the long exposure uncertainties clearly differ between the different cases. The very different low-frequency 
behavior of the 4 cases 
mostly goes away once one averages over all wind directions, as seen in \fref{figDualStarRotAvgPS}. 
The flat power spectrum at low frequencies means that astrometric uncertainty goes as the conventional $\Delta T^{-1/2}$, so long as the integration time $\Delta T$ is significantly greater than the wind crossing time of the baseline. 
Note also that higher wind velocities {\em decrease} the long-exposure uncertainties (\ie, the low frequency power), because a larger number of independent wavefront samples are included in the average. 
The long exposure uncertainty in dual-star astrometry comes from integrating the product of this frequency spectrum with a sinc$^2$ function, which is approximately the same as scaling the zero frequency intercept by $\sqrt{\Delta T}$, \ie:
\begin{align}
\sigma_\theta(\Delta T) &= \sqrt{2 \int_0^{f_c} \Theta (f) {\rm sinc}^2(\pi f \Delta T) df }\\
 &\approx  \Theta (0) \Delta T^{-1/2}.
\end{align}

\begin{figure}
\centering
\includegraphics[width=0.9\textwidth]{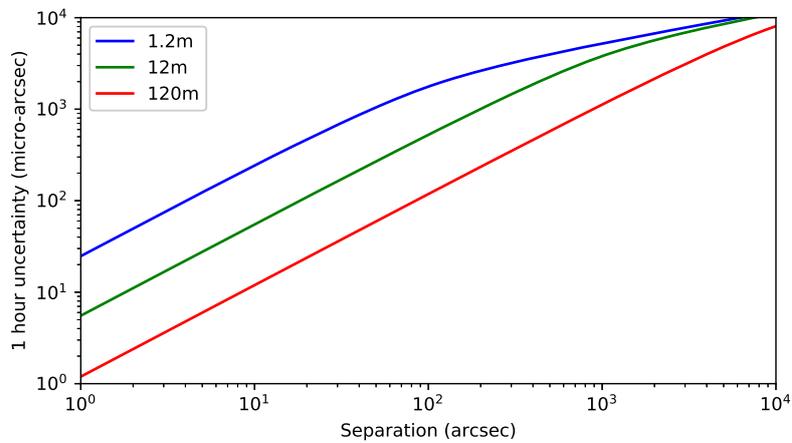}
\caption{Astrometric uncertainty as a function of separation in the case of dual-star astrometry, with the same parameters as \fref{figDualStarPS}. See electronic edition for a color version of this figure.}
\label{figErrorSeparation}
\end{figure}

This result is shown in \fref{figErrorSeparation} for the same atmospheric conditions as \fref{figDualStarPS}, for binary separation parallel to the baseline, and for the same baselines as discussed in Shao and Colavita (1992) \cite{Shao&Colavita1992}. Different assumptions on turbulence profiles slightly change the numerical values in this relationship, but the key asymptotic relations remain the same. These are:
\begin{align}
\sigma_\theta \propto | B^{-2/3}\boldsymbol{\theta}| {\rm~for~}|\boldsymbol{\theta}| \ll |\boldsymbol{B}|/\bar{h}, \Delta T \gg |\boldsymbol{B}|/|\boldsymbol{\bar{v}}|\\
\sigma_\theta \propto |\boldsymbol{\theta}|^{1/3} {\rm~for~}|\boldsymbol{\theta}| \gg |\boldsymbol{B}|/\bar{h}, \Delta T \gg  \bar{h} |\boldsymbol{\theta}|/|\boldsymbol{\bar{v}}.
\end{align}

\subsection{Multiple Star Astrometry}
\label{SSec:MultipleStarAstrometry}

In practice, the great advantage of past astrometric imaging instruments over long baseline interferometry has been the ability for multiple reference stars to be observed. By tying together many astrometric fields with stars in common, multiple-star astrometry in principle means that an astrometric grid over the whole sky can be constructed. In practice, this is best done from space, with HIPPARCOS and Gaia providing the reference star grids for the future.  
Ground-based and future space-based narrow angle astrometry will be tied to these
grids and have relative precision within a field limited by Gaia (uncertainties as small as one part in $10^{10}$), but in principle can obtain much better precisions than Gaia for individual stars.

Multiple-reference star astrometry has never been attempted in long baseline interferometry (dual star astrometry is difficult enough!) but has been used in precision astrometric experiments on single telescopes using adaptive optics \cite{Lazorenko&Lazorenko2004,Cameron+2009}. Although calibration has been difficult, this use of multiple reference stars has enabled astrometry at the 100 micro-arcsecond level, even on the relatively short baselines afforded by single telescopes.

We will consider idealized narrow-angle astrometry for two classes of multiple reference star astrometry here: two reference stars either side of a target, and three reference stars around a target with arbitrary vector separations. The angular separation vectors of these stars are $\boldsymbol{\theta}_1$, $\boldsymbol{\theta}_2$ and $\boldsymbol{\theta}_3$. 

In the linear (2-star) case, where the reference stars are on either side of the target,
the differential piston astrometric signal is formed from the fringe phases as:
\begin{align}
\frac{\Delta \theta \lambda}{2\pi} &= |B| \Big( [ \varphi(\mathbf{B}) - \varphi(0) ] - \nonumber \\
	& \frac{|\theta_2|}{|\theta_1| + |\theta_2|}[ \varphi(\mathbf{B} + \boldsymbol{\theta}_1\bar{h}) - \varphi(\boldsymbol{\theta}_1\bar{h})]  -
	\frac{|\theta_1|}{|\theta_1| + |\theta_2|}[ \varphi(\mathbf{B} + \boldsymbol{\theta}_2\bar{h}) - \varphi(\boldsymbol{\theta}_2\bar{h})] \Big).
\label{eqnPhaseDoubleDifference}
\end{align}
The spatial power spectrum for the symmetric two reference star case is given by:
\begin{align}
\Theta (\boldsymbol{\kappa}) &= 0.00928 |B|^{-2} \lambda^2 r_0^{-5/3} |\boldsymbol{\kappa}|^{-11/3} \sin^2(\pi\mathbf{B} \cdot \boldsymbol{\kappa}) \sin^4(\pi \bar{h}\boldsymbol{\theta} \cdot \boldsymbol{\kappa}).
\end{align}
The only difference to the case with one reference star is that the second sine function is raised to the fourth power. 

For the three star case, the differential piston astrometric signal is formed from:
\begin{align}
\frac{\Delta \theta \lambda}{2\pi} &= |B| \Big( [ \varphi(\mathbf{B}) - \varphi(0) ] - 
	\Sigma_{i=1}^3 w_i  [ \varphi(\mathbf{B} + \boldsymbol{\theta}_i \bar{h}) - \varphi(\boldsymbol{\theta}_i \bar{h})] \Big).
\label{eqnFourStar}
\end{align}
The weights, $\boldsymbol{w} = \{w_1, w_2, w_3 \}$, come from solving the $3 \times 3$ linear system,
\begin{align}
[ \boldsymbol{\theta}_1 \boldsymbol{\theta}_2 \boldsymbol{\theta}_3 ] \cdot \boldsymbol{w} &= 0 \\
\Sigma_{i=1}^3 w_i = 1 \, ,
\end{align}
%
and ensure that the lowest spatial frequency aberrations in the upper atmosphere turbulence cancel out, with the higher spatial frequency terms averaging out relatively quickly.  
The power spectrum then becomes:
\begin{align}
\Theta (\boldsymbol{\kappa}) &= 0.00232 |B|^{-2} \lambda^2 r_0^{-5/3} |\boldsymbol{\kappa}|^{-11/3} \sin^2(\pi\mathbf{B} \cdot \boldsymbol{\kappa}) 
| 1- \Sigma_{i=1}^3 w_i \exp(i 2 \pi \bar{h}\boldsymbol{\theta}_i \cdot \boldsymbol{\kappa})|^2.
\label{eqnFourStarFourier}
\end{align}

\begin{figure}
\centering
\includegraphics[width=0.8\textwidth]{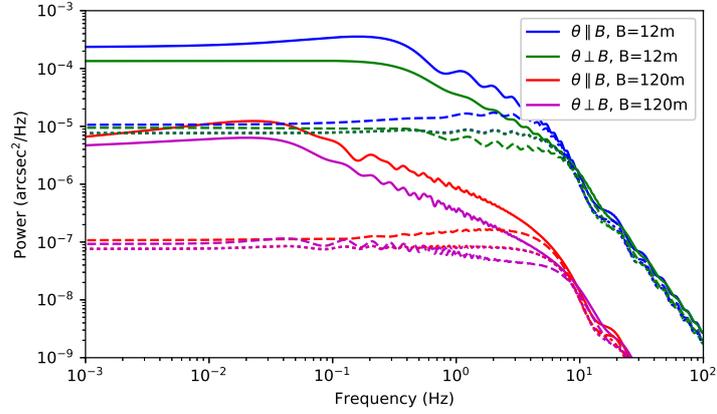}
\caption{Temporal power spectra in the case of one reference star (solid), two symmetrically placed reference stars (dashed) and three
symmetrically placed reference stars (dotted). 
Low frequency power is reduced by a factor of $\sim$100 for the 120\,m baseline case by the 
additional reference star(s), decreasing required integration times by the same factor. See electronic edition for a color version of this figure.}
\label{figThreeStarPS}
\end{figure}

\begin{figure}
\centering
\includegraphics[width=0.8\textwidth]{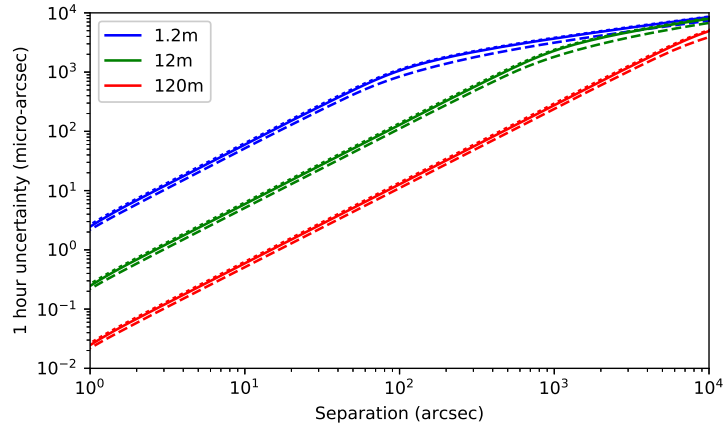}
\caption{Astrometric uncertainty as a function of separation in the case of astrometry with three reference stars. The solid line is for equal spacing (at \SI{120}{\degree} angles), the dashed line is for spacings of 0.5, 1 and 1.5 times the mean separation, and the dotted line is for equal separations, but \SI{100}{\degree}, \SI{100}{\degree} and \SI{160}{\degree} angular spacing of reference stars. See electronic edition for a color version of this figure.}
\label{figErrorSeparationFour}
\end{figure}

The power spectra for the 2 and 3 reference star cases are shown in \fref{figThreeStarPS}, and the angular- and baseline-dependence of the uncertainty of the 3-star case shown in 
\fref{figErrorSeparationFour}. The two reference star uncertainty is nearly indistinguishable from the case of three reference stars where two are separated by 160 degrees, so is not plotted.
With these additional reference stars, the astrometric uncertainty is significantly smaller at small angular separations, with an an asymptotic power law with uncertainty proportional to $|\boldsymbol{B}|$ and $|\boldsymbol{\theta}|^{4/3}$.


\section{Limitations of Real Instruments}
\label{Sec:NonCommonPath}

\subsection{A Narrow Angle Interferometer Archetype}
\label{SSec:NarrowAngleArchetype}

\begin{figure}
\centering
\def\svgwidth{0.7\textwidth}
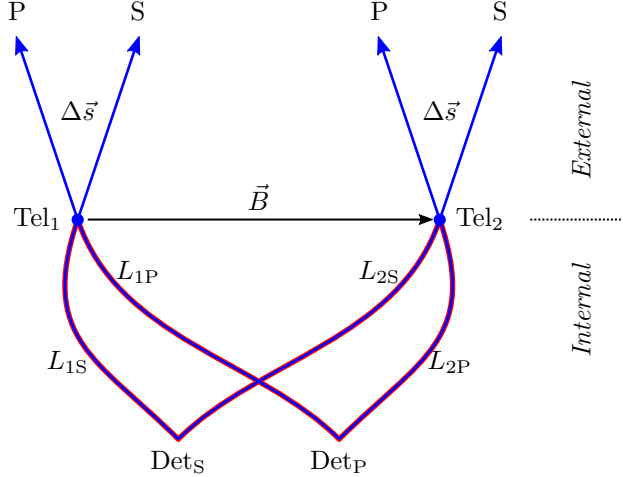
\caption{A narrow angle interferometer archetype. Two independent interferometric detectors $\mathrm{Det_P}$ and $\mathrm{Det_S}$ observe two targets $\mathrm{P}$ and $\mathrm{S}$ with two telescopes $\mathrm{Tel}_1$ and $\mathrm{Tel}_2$. A metrology system (in red) measures the differential internal optical path $\Delta{L}_{int}=(L_{2,P}-L_{1,P})-(L_{2,S}-L_{1,S})$.}
\label{Fig:NarrowAngleAstrometryConcept}
\end{figure}

For sky coverage reasons, astrometric observations are carried out between targets with separations larger than the diffraction limit of a single telescope.
Since the field of view of interferometric instruments is traditionally the diffraction limit of its telescopes, an astrometric observation requires independent interferometric detectors, one per observed target.
As for single telescopes where the astrometric measurement usually amounts to counting the pixel distance between objects in the focal plane, the distance between interferometric detectors needs to be measured.
This is taken care of by a metrology system, as conceptually illustrated in \fref{Fig:NarrowAngleAstrometryConcept} for a two targets observation.
This metrology measures the internal optical path double difference $\Delta{L}_{int}$ between the two targets ($P$, $S$) and the two 
telescopes ($1$, $2$):
\begin{equation}
    \Delta{L}_{int}=(L_{2,P}-L_{1,P})-(L_{2,S}-L_{1,S}) \, .
\end{equation}
When the delay lines equalize the optical path through the telescopes for each target such that $\Delta{L}_{int} + \Delta{L}_{ext} = 0$, the differential external delay $\Delta{L}_{ext} = - \vec{B} \cdot \Delta\vec{s}$ is directly given by the internal metrology measurement:
\begin{equation}
    \Delta{L}_{int} = \vec{B} \cdot \Delta{\vec{s}}.
    \label{Eq:AstrometricObservation}
\end{equation}
Knowing the baseline $\vec{B}$, and measuring the differential internal optical path $\Delta{L}_{int}$ with the metrology, the separation $\Delta\vec{s}$ projected on the baseline direction is determined.
For a complete knowledge of the separation vector $\Delta{\vec{s}}$, multiple baseline orientations,  super-synthesis, or a combination of both are needed.

In practice, the metrology system measures the differential internal optical path $\Delta{L_{int}}$ to within an zero-point $\Delta{L}_{int,0}$, determined at the same time as the separation vector $\Delta{\vec{s}}$:
\begin{equation}
    \Delta{L}_{int} - \Delta{L}_{int,0} = \vec{B} \cdot \Delta{\vec{s}}.
    \label{Eq:AstrometricObservationWithZero}
\end{equation}
One of the two following methods is used to independently identify the zero-point and the separation.
Either the astrometric observation is performed over a range of hour angles  large enough for the sine-like $\vec{B}\cdot\Delta\vec{s}$ to separate from the constant $\Delta{L}_{int,0}$ zero-point, or the separation vector $\Delta\vec{s}$ is inverted or zeroed to independently determine the metrology zero-point.
An inversion of the separation is achieved by swapping the two observed targets between the two detectors, whereas a zeroing is obtained with one single target observed simultaneously by the two detectors.
Both operations require special capabilities at the level of the dual-star, but for simplicity and efficiency reasons, the swap approach has always been favored.

\subsection{Back-of-the-envelope error budget}
\label{SSec:ErrorBudget}

In order to develop a realistic error budget for a narrow-angle astrometric observation, \eref{Eq:AstrometricObservation} needs to be supplemented by the effects of atmospheric turbulence and detector photon noise:
\begin{equation}
  \Delta{L}_{atm} + \Delta{L}_{int} + \Delta{L}_{det} = \vec{B} \cdot \Delta\vec{s} \, .
  \label{Eq:Astrometry}
\end{equation}
Since the contributions of the photon noise and the atmosphere, both with zero mean, have been extensively covered in  \sref{Sec:Photon} and \sref{Sec:Atmosphere}, respectively, they will not be considered any further in this section.

Neglecting the geometry of the observation, the error on the separation $\delta\Delta{s}$ relates to the error on the baseline knowledge $\delta{B}$ and on the internal optical path difference $\delta\Delta{L_{int}}$ as follows:
\begin{equation}
  \left( \frac{\delta\Delta{s}}{\Delta{s}} \right)^2 = \left( \frac{\delta{B}}{B} \right)^2 + \left( \frac{\delta\Delta{L_{int}}}{\Delta{L_{int}}} \right)^2 \, .
  \label{Eq:ErrorBudget}
\end{equation}
The fractional error on the separation, $\delta\Delta{s}/\Delta{s}$, is a combination of the fractional error on the baseline $\delta{B}/B$ and the fractional error on the differential internal optical path difference $\delta\Delta{L_{int}}/\Delta{L_{int}}$.
To illustrate, a precision/accuracy of 10~micro-arcseconds for a \SI{10}{\arcsecond} separation requires better than a $10^{-6}$ fractional error on both the baseline and the differential internal optical path difference.
This corresponds to better than \SI{\sim100}{\micro\meter} on the baseline, and better than \SI{\sim5}{\nano\meter} on the differential internal optical path difference.
Since the differential internal optical path difference is deduced from the differential phase $\Delta\phi_{met}$ of the metrology laser, as
\begin{equation}
  \Delta{L_{int}} = \frac{\lambda_{met}}{2\pi}\Delta\phi_{met} \, ,
  \label{Eq:MetrologyPhaseToOpticalPath}
\end{equation}
the knowledge of the laser wavelength $\lambda_{met}$ is critical.
To obtain a fractional error of $10^{-6}$ on the separation, an equivalent fractional precision/accuracy is required on the metrology wavelength.

Depending on the scientific objectives of the astrometric observations, an absolute measurement of the separation might not be necessary, \ie, it might be sufficient to know the baseline length and the metrology wavelength to a scaling factor.
In this case, the measured separation will be affected by that same scaling factor.
However, since astrometry is a technique to observe the universe in motion, an eventual scaling factor has to be at least constant, at the levels described above, over the multi-year duration of the measurement.

\subsection{Non-common path}
\label{SSec:NonCommonPath}

The real challenge of interferometric astrometry is making sure that the metrology measures exactly the optical path experienced by the astronomical light.
Any mismatch between the two qualifies as a non-common path error.
Detailed analysis for some of these non-common path errors can be found in the literature\cite{Colavita2009}; we discuss some of them below.

\textbf{Path coverage}:
Ideally, the astrometric metrology should measure the internal differential optical path difference from the beam combination point to the primary space where the baseline is defined.
In practice, depending on the metrology implementation this may not be completely achievable.
A telescope-differential metrology (measuring $L_{2,P}-L_{1,P}$ and $L_{2,S}-L_{1,S}$) injected on the back side of the beam combiner might not be able to go all the way up to the telescope, being blocked by the phase errors induced by the deformable mirror of an adaptive optics system (\eg, ASTRA on Keck Interferometer).
On the other hand, a target-differential system (measuring $L_{1,P}-L_{1,S}$ and $L_{2,P}-L_{2,S}$) would be able to go through the deformable mirror 
allowing the internal differential optical path difference to be 
 measured in primary space (\eg, Gravity on the VLTI), but at the expense of a non-common path between the injection on the back side of the two beam combiners.

\textbf{Beam walk}:
When the footprints of the stellar and metrology lights are not the same, either due to a smaller metrology beam propagating in the central obscuration of the telescope (\eg, PRIMA on the VLTI), or to a smaller metrology sensor (\eg, Gravity on the VLTI), and when this footprint moves on imperfect optical elements, the stellar and metrology lights do not experience the same optical path.
This effect can either appear as an additional noise in the astrometric measurement when, \eg,  internal turbulence induces the beam motion, or as a bias when the motion is correlated with the astrometric observation sequence.
  
\textbf{Chromatic effects}:
All interferometers have used the metrology laser at a wavelength outside the stellar bandpass.
The refractive index of the material used in transmission has an impact on the measurement of the internal optical path.
When the differential optical delay is not implemented in vacuum (\eg, in fluoride glass fibers for Gravity on the VLTI), \eref{Eq:MetrologyPhaseToOpticalPath} needs to be modified to include the refractive index $n$ of the material.
\begin{equation}
  \Delta{L_{int}} = \frac{n_{sci}}{n_{met}}\frac{\lambda_{met}}{2\pi}\Delta\phi_{met}
\end{equation}
The same level of requirement on the metrology wavelength applies to the refractive index.

\textbf{Polarization}:
Having all the arms of an interferometer with identical polarization properties is sufficient only to observe unpolarized sources.\cite{Buscher+2009}
The astrometric metrology being built around highly polarized lasers, these systems are bound to have issues that can impact astrometric measurements.
As an illustration, the vibration metrology of the Keck Interferometer needed its polarization state to be adjusted with the attitude of the telescopes to operate reliably\cite{Colavita+2013}, and the field derotator of the VLTI auxiliary telescopes had to be polarization-compensated to extend the astrometric metrology of PRIMA to the secondary mirror of the telescopes\cite{Woillez+2014b}.
Due to the polarization properties of SgrA*, the impact of polarization on astrometry has been studied for the Gravity instrument of VLTI\cite{Lazareff+2014}.
However, polarization has not been identified (yet) as the limiting factor of the astrometric performance of the instrument.

The non-common path terms above are the main reason for choosing the separation swap over the hour-angle coverage method of determining the metrology zero-point (see \sref{SSec:NarrowAngleArchetype}).
Regular swaps on short timescales help reject temporal variations of non-common path effects into the metrology zero-point, reducing the impact on the separation measurement.
This approach works as long as the swap operation itself does not introduce an astrometric bias, by, \eg, impacting the primary-space conjugation of the narrow-angle baseline\cite{Sahlmann+2013}, as described in the next section.

\subsection{Astrometric Baselines}
\label{SSec:AstrometricBaselines}

In practice, one of the major limitations in astrometric interferometry is the definition of an narrow-angle astrometric baseline\cite{Woillez&Lacour2013}. Where three or more reference stars are used, as is the case in single-telescope astrometry, variations in the astrometric baseline in principle cancel out by construction (\eref{eqnFourStar}) and it simply becomes a scaling factor for astrometric shifts. However, for only a single reference star (\ie, all existing or planned long-baseline instruments), the astrometric baseline uncertainty becomes a relative astrometric uncertainty, as shown in \eref{Eq:ErrorBudget}.

The definition of the narrow-angle astrometric baseline comes from the requirement of no optical path gap between the coverage of the internal metrology measurement, $\Delta{L_{int}}$, and the external delay, $\vec{B}\cdot\vec{\Delta{s}}$.
Said differently, the internal metrology has to reach, at each telescope, the points that define the narrow-angle baseline $\vec{B}$.
Tying the baseline vector $\vec{B}$ to the Earth reference frame (ITRS), the transformation of the baseline to the star reference frame (ICRS) is sufficient to determine the $\vec{B}\cdot\Delta\vec{s}$ scalar product.
The baseline vector must therefore be expressed in primary space, \ie, before any conjugation by the primary mirrors of the telescopes.
If the metrology end-point does not reach primary space, its conjugation to primary space needs to be monitored against a primary space reference.
Most larger telescope employed so far for astrometric interferometry have a pupil with a central obscuration caused by the secondary mirror.
As such, the telescope pivot point, 
the only Earth-fixed point of an ideal telescope that the pointing axis intersects,
cannot be observed from inside the instrument and therefore reached by astrometric metrology.
To overcome this issue, different methods have been considered.
For the ASTRA project on the Keck Interferometer (see \sref{SSec:KeckInterferometer}), the metrology was terminated inside the adaptive optics system and monitored with respect to a primary space reference inserted  between the segments of the primary mirror.
For the Gravity instrument on the VLTI (see \sref{SSec:VLTI_Gravity}), the metrology extended directly into primary space, up to metrology receivers located on the spiders supporting the secondary mirror of the telescopes.

An absolute knowledge of the baseline is necessary for an absolute measurement of the separation.
This raises the issue of the narrow-angle baseline absolute calibration.
Past and present interferometers do not currently have readily available binary targets with separations known at the level of a few micro-arcseconds\footnote{This might change soon with the upcoming data release of the Gaia mission\cite{GaiaCollaboration+2016}.} and bright enough to be observable.
This has been circumvented by transferring the wide-angle baseline, measured with a set of single stars with known positions and large separations, to the narrow angle baseline\cite{Hrynevych+2004,Colavita2009,Woillez+2010,Sahlmann+2013}.
This transfer adds a second layer of complexity to the baseline issue, as the adjustment  of a wide-angle baseline model alone is not sufficient for the transfer from wide-angle to narrow-angle to be under control.
The wide-angle pivot point and the absolute internal optical path difference, $L_{int}$,  need to be controlled as well for the single target measurements\cite{Woillez&Lacour2013}.

\section{Past, Current and Planned Instruments}
\label{Sec:PastPresentFuture}

\subsection{Mark III}
\label{SSec:MarkIII}

On Mount Wilson (US), the Mark III \cite{Shao+1988} was the first interferometer to carry out a dual star observation \cite{Colavita1994}, following the seminal paper of Shao \& Colavita (1992) \cite{Shao&Colavita1992}.
The focal plane of the interferometer was modified to simultaneously observe the two components of the double star $\alpha$ Gem, and study the atmospheric turbulence in the narrow-angle regime.
No disagreement was found between this first observation and the predictions of a Kolmogorov turbulence model.

\subsection{Palomar Testbed Interferometer}
\label{SSec:PTI}

A more systematic exploration of the parameter space required a dedicated dual-star interferometer.
The successor of the Mark III, the Palomar Testbed Interferometer (PTI)\cite{Colavita+1999} on Mount Palomar (US), was constructed to verify the atmospheric limits and demonstrate the technologies needed for narrow angle astrometry.
PTI had three \SI{40}{\centi\meter} siderostats equipped with a dual-star separator generating a beam 
for each of the two observed targets, dual delay lines and differential delay lines to equalize the optical path for both targets, two interferometric detectors, one of them acting as a fringe tracker to compensate the atmospheric perturbations, and an astrometric metrology system covering the full optical path from the interferometric detectors to a retro-reflector at each siderostat in the path common to the two targets.
The first narrow-angle astrometric measurements were obtained on the \SI{\sim31.5}{\arcsecond} separation binary 61 Cyg \cite{Shao+1999,Lane+2000}, and reached a stability in the range 
100 -- 170 micro-arcseconds,
which corresponds to a fractional error of $\sim5\times10^{-5}$.
As a demonstrator with small apertures, PTI had limited sensitivity, preventing further astrometric investigations.

PTI also carried out very-narrow-angle astrometric observations, under a project named PHASES\cite{Muterspaugh+2010}, where the target pairs were fully contained inside the \SI{\sim1}{\arcsecond} diffraction limit of its \SI{40}{\centi\meter} siderostats, and therefore did not require a dual-star module.
A beamsplitter  separated the light between a fringe tracker compensating the atmospheric piston and a science camera scanning through the two fringe packets generated by the pair.
The astrometric separation was deduced from the separation within a scan between the two fringe packets.
This type of astrometric observations, in the double packet regime\cite{Lachaume&Berger2013}, is fundamentally different in concept and implementation from the narrow-angle astrometry presented in this review, being more related to imaging astrometry\cite{Woillez&Lacour2013}.

\subsection{Keck Interferometer}
\label{SSec:KeckInterferometer}

Successor of PTI, the Keck Interferometer project\cite{Colavita+2013} on Mauna Kea (US) started receiving funding in 1998\cite{Colavita+1998}.
It was planned as an extension of the Keck Observatory twin \SI{10}{\meter} telescopes, with four additional \SI{1.8}{\meter} telescopes\cite{Bell+2004}.
The \SI{1.8}{\meter} telescopes had been planned from the beginning for narrow-angle astrometry, including a dual-star capability and some equipment to monitor the telescope's pivot points and the transfer of the narrow-angle baseline onto the wide-angle baseline\cite{Hrynevych+2004}.
The delay lines, similar to the ones at PTI, also had a dual-star capability.
Despite having been built, the astrometric project with the \SI{1.8}{\meter} telescopes was cancelled in 2006.

In 2006, the ASTRA project\cite{Pott+2009} on the Keck Interferometer started implementing a phase referencing and narrow-angle astrometry capability, but this time on the main \SI{10}{\meter} Keck telescopes, with the observation of the galactic center as a main objective.
ASTRA developed\cite{Woillez+2010} a dual-star capability at the focus of the adaptive optics systems equipping the two \SI{10}{\meter} telescopes, a \SI{1319}{\nano\meter} double-pass astrometric metrology covering the optical path from the two existing fringe trackers to common retro-reflectors located in front of the deformable mirrors of each adaptive optics system, and a camera monitoring the conjugation of the astrometric baseline to primary space.
Unfortunately, after delivering an off-axis fringe tracking capability\cite{Woillez+2014a}, ASTRA did not have time to demonstrate any astrometric performance before the Keck Interferometer ceased operation\cite{Ragland+2012}.

\subsection{Sydney University Stellar Interferometer - MUSCA/PAVO}
\label{SSec:SUSI}

The Sydney University Stellar Interferometer (SUSI)  attempted very-narrow-angle dual-star metrology over the years 2011 to 2013 \cite{Kok13}, in order to explore limits for astrometrically detecting planets in binary star systems. The dual-star system MUSCA (Micro-arcsecond University of Sydney Companion Astrometry) measured the fringes on a uniaxial beam combiner over the wavelength range 0.77--0.9\,$\mu$m on either the primary or the secondary star, while the companion beam combiner PAVO tracked and recorded telescope-differential optical path differences on the primary star over the wavelength range 0.54--0.76\,$\mu$m. Operating only over a field of view of up to $\sim$3\,arcseconds and with a magnification of only a factor of 3 through the delay lines, optical paths were assumed equal for both observed stars stars up to the MUSCA/PAVO dichroic split, with a target-differential metrology system between the beam combiners. The wide-angle baseline and imaging baselines were kept in sync at the cm level (with upgrades identified for mm-level accuracy) by a pupil viewing and alignment system. 

This system successfully demonstrated phase-referenced interferometry at the level of 100\,$\mu$as in the most favorable conditions. Potential paths to lower astrometric uncertainty were identified, including a more extensive control of non-common path aberrations and removing spurious reflections of metrology signals. However, the sensitivity of PAVO as a fringe tracker was ultimately inadequate for the exoplanet science case (approximately V$\sim$5 in good conditions).

\subsection{Very Large Telescope Interferometer - PRIMA}
\label{SSec:VLTI_PRIMA}

In Europe and Chile, the astrometry and phase-referenced imaging project PRIMA\cite{Delplancke2008} of the Very Large Telescope Interferometer started being developed in 2000.
It provided a dual-star capability\cite{Delplancke+2004} for the \SI{1.8}{\meter} Auxiliary Telescopes (AT) and \SI{8}{\meter} Unit Telescopes (UT), vacuum differential delay lines\cite{Pepe+2008}, a pair of two-telescopes fringe trackers\cite{Sahlmann+2009}, and an astrometric metrology system\cite{Leveque+2003}.
The installation started in 2008 and the first astrometric observations were carried out in 2011.
By mid-2012, it became apparent that the astrometric performance of 
3~milli-arcseconds for a \SI{10}{\arcsecond} separation, which corresponded to a fractional error of $\sim3\times10^{-4}$, was not meeting the $10^{-5} \sim 10^{-6}$ expectation\cite{Sahlmann+2013}.
Major shortcomings were identified in the implementation of the fringe sensors, the star separators, and the astrometric metrology defining the baseline.
Polarization effects introduced by the field derotators of the star separators, combined with a polarization-based fringe sensor design, impacted fringe tracking performance and sensitivity, while the termination of the astrometric metrology inside the star separators induced large primary-space-conjugated astrometric baseline motions.
These issues were partly addressed by compensating the polarization properties of the field derotators and extending the metrology to the secondary mirror of the telescopes.
In its improved configuration, PRIMA demonstrated\cite{Woillez+2014b} a performance of 
800~micro-arcseconds for the same \SI{10}{\arcsecond} separation, but with a very unfavorable projected baseline of \SI{30}{\meter}.
Extrapolated to \SI{150}{\meter}, this amounted to 
160~micro-arcseconds, or a fractional error of $1.6\times10^{-5}$, still short of the requirements of an exoplanet detection and characterization campaign\cite{Launhardt+2008}.
Unfortunately, in direct competition with Gaia\cite{GaiaCollaboration+2016}, the space astrometry mission of ESA, the PRIMA project was put on hold in 2014 and cancelled the following year.

\subsection{Very Large Telescope Interferometer - Gravity}
\label{SSec:VLTI_Gravity}

Currently, the only narrow-angle astrometry instrument in operation on any interferometer is Gravity\cite{Eisenhauer+2011} on the VLTI.
Started in 2008, Gravity is a K-band fully cryogenic two-object beam combiner installed at the VLTI focus.
Its main objective is the study of the supermassive black hole at the center of our galaxy.
As a second generation VLTI instrument, Gravity benefited from the experience with the previous generation of interferometric instruments, of which the evolution of its astrometric metrology design\citet{Gillessen+2010,Gillessen+2012} is a good illustration.
It includes two fiber-fed integrated optics 4-telescope beam combiners\cite{Jocou+2014}, a low-noise infrared detector for its fringe tracker, active control loops for the stabilization of the fields and pupils\cite{Anugu+2016}, and a novel \SI{1908}{\nano\meter} astrometric metrology system\cite{Lippa+2016}.
The dual-star capability is contained inside the instrument itself and therefore has to work within the field of view delivered by the VLTI, \SI{\sim2}{\arcsecond} for the UTs and \SI{\sim4}{\arcsecond} for the ATs, which limits the sky coverage compared to past instruments but also reduces non-common path effects between the two observed objects.
The astrometric metrology design chosen for Gravity is unique.
The metrology laser light is split into two differentially phase-shifted beams, back-injected into the integrated optics combiners, and propagated toward the telescopes.
A local metrology pickup measures the optical path difference between the two objects at each telescope input of the instrument.
The signal level there is high enough to record the variations of the optical path differences, without wrapping errors, continuously throughout an astrometric observation.
Because the metrology measurement is target-differential, rather than telescope-differential, the 
Gravity metrology can propagate past the deformable mirrors of the adaptive optics systems.
Four additional metrology receivers are installed on the spiders of each telescope; they directly materialize the astrometric baseline in primary space.
The pupil control loops measure the relative positions between the metrology receivers inside the instrument and the ones on the telescopes, and help unwrap the telescope metrology signals between subsequent measurements of an astrometric observation sequence.
Because the differential delay capability is implemented in optical fibers, the control of dispersion is critical to the performance of the instrument, as presented in its error budget\cite{Lacour+2014}.
The first observations with Gravity, including its astrometric mode, were carried out in $2015\sim2016$\cite{GravityCollaboration+2017}.
As of late 2017, astrometric observations are carried out with a combination of the optical path measurements by the internal metrology pickup and the corrections from the pupil control loop.
The fractional performance achieved is on the order of $5\times10^{-5}$.
The extension of the metrology to the telescope receivers, which was still under commissioning, should improve this fractional performance.

\section{Conclusions and Forward Look}
\label{Sec:Conclusions}

This chapter has focused on the motivation, principles and practice of ground-based astrometric interferometry, demonstrating that 
$\sim 1$~micro-arcsecond precision is possible, but the technique still has many challenges, as the many attempts have shown. Recent experience has demonstrated that for sufficiently large telescopes, astrometric interferometry can observe astrophysically interesting and faint targets. The Gravity instrument for the VLTI has demonstrated routine astrometric precisions at the level of 
$\sim 50$~micro-arcseconds\cite{Martina18}, and phase-referenced imaging down to a magnitude $m_K$  fainter than 17\cite{Widmann18}. The astrometric precision of Gravity for the primary black hole science case \cite{Gravity18} is enabled by the ability to track on fainter stars (\eg, Sgr A* itself), in turn enabled by off-axis fringe tracking and phase referencing, which is in turn enabled by the large apertures of the VLTI's Unit Telescopes. Indeed, the fundamental uncertainties described in \sref{Sec:Atmosphere} are as applicable to phase-referenced imaging as they are to dual-star or multi-star astrometry.

For ground-based instrumentation with large apertures, one of the key limiting effects, not described here, is the isoplanatic angle for natural guide star adaptive optics. Achieving  sufficient sky coverage for a significantly expanded impact with a ground-based system then requires laser guide stars, and possibly a multi-conjugate system at each telescope to increase the field of view in \SI{8}{\meter} class telescopes to  \SI{>30}{\arcsecond}.  However, this potentially adds systematic uncertainties \cite{Neichel+2014}, and therefore almost certainly requires 3 reference stars. Such a system would be a major undertaking, but as shown in \sref{Sec:Photon} and~\sref{Sec:Atmosphere}, it could achieve a precision level of less than 
$\sim 1$~micro-arcsecond and be limited primarily by the Earth's atmosphere.

There are no major space-based interferometer studies at the present time, with the Space Interferometer Mission (SIM) \cite{Unwin08} having been cancelled, and the cost of a 
competitive mission based on those previous studies being in the $>$1\,B\$ range. Nonetheless, in the context of Gaia's anticipated exoplanet results, it may be worthwhile reconsidering a range of space interferometer concepts in the near future.





\bibliographystyle{ws-rv-van}
\bibliography{astrometric-interferometry}

\begin{thebibliography}{62}
\providecommand{\natexlab}[1]{#1}
\providecommand{\url}[1]{\texttt{#1}}
\expandafter\ifx\csname urlstyle\endcsname\relax
  \providecommand{\doi}[1]{doi: #1}\else
  \providecommand{\doi}{doi: \begingroup \urlstyle{rm}\Url}\fi

\bibitem{Perryman2012}
M.~{Perryman}, {The history of astrometry}, \emph{European Physical Journal H}.
  {\bf 37}, \penalty0 745--792 (Oct., 2012).
\newblock \doi{10.1140/epjh/e2012-30039-4}.

\bibitem{Dyson+1920}
F.~W. {Dyson}, A.~S. {Eddington}, and C.~{Davidson}, {A Determination of the
  Deflection of Light by the Sun's Gravitational Field, from Observations Made
  at the Total Eclipse of May 29, 1919}, \emph{Philosophical Transactions of
  the Royal Society of London Series A}. {\bf 220}, \penalty0 291--333,
  (1920).
\newblock \doi{10.1098/rsta.1920.0009}.

\bibitem{vandeKamp1963}
P.~{van de Kamp}, {Astrometric study of Barnard's star from plates taken with
  the 24-inch Sproul refractor.}, \emph{\aj}. {\bf 68}, \penalty0 515--521
  (Sept., 1963).
\newblock \doi{10.1086/109001}.

\bibitem{Pravdo&Shaklan2009}
S.~H. {Pravdo} and S.~B. {Shaklan}, {An ultracool Star's Candidate Planet},
  \emph{\apj}. {\bf 700}, \penalty0 623--632 (July, 2009).
\newblock \doi{10.1088/0004-637X/700/1/623}.

\bibitem{Bean+2010}
J.~L. {Bean}, A.~{Seifahrt}, H.~{Hartman}, H.~{Nilsson}, A.~{Reiners},
  S.~{Dreizler}, T.~J. {Henry}, and G.~{Wiedemann}, {The Proposed Giant Planet
  Orbiting VB 10 Does Not Exist}, \emph{\apjl}. {\bf 711}, \penalty0 L19--L23
  (Mar., 2010).
\newblock \doi{10.1088/2041-8205/711/1/L19}.

\bibitem{Liu+2016}
M.~C. {Liu}, T.~J. {Dupuy}, and K.~N. {Allers}, {The Hawaii Infrared Parallax
  Program. II. Young Ultracool Field Dwarfs}, \emph{\apj}. 833:\penalty0 96
  (Dec., 2016).
\newblock \doi{10.3847/1538-4357/833/1/96}.

\bibitem{Perryman+2001}
M.~A.~C. {Perryman}, K.~S. {de Boer}, G.~{Gilmore}, E.~{H{\o}g}, M.~G.
  {Lattanzi}, L.~{Lindegren}, X.~{Luri}, F.~{Mignard}, O.~{Pace}, and P.~T. {de
  Zeeuw}, {GAIA: Composition, formation and evolution of the Galaxy},
  \emph{\aap}. {\bf 369}, \penalty0 339--363 (Apr., 2001).
\newblock \doi{10.1051/0004-6361:20010085}.

\bibitem{Perryman+2014}
M.~{Perryman}, J.~{Hartman}, G.~{\'A}. {Bakos}, and L.~{Lindegren},
  {Astrometric Exoplanet Detection with Gaia}, \emph{\apj}. 797:\penalty0 14
  (Dec., 2014).
\newblock \doi{10.1088/0004-637X/797/1/14}.

\bibitem{GravityCollaboration+2017}
{Gravity Collaboration}, R.~{Abuter}, M.~{Accardo}, A.~{Amorim}, N.~{Anugu},
  G.~{{\'A}vila}, N.~{Azouaoui}, M.~{Benisty}, J.~P. {Berger}, N.~{Blind},
  H.~{Bonnet}, P.~{Bourget}, W.~{Brandner}, R.~{Brast}, A.~{Buron},
  L.~{Burtscher}, F.~{Cassaing}, F.~{Chapron}, {\'E}.~{Choquet},
  Y.~{Cl{\'e}net}, C.~{Collin}, V.~{Coud{\'e} Du Foresto}, W.~{de Wit}, P.~T.
  {de Zeeuw}, C.~{Deen}, F.~{Delplancke-Str{\"o}bele}, R.~{Dembet}, F.~{Derie},
  J.~{Dexter}, G.~{Duvert}, M.~{Ebert}, A.~{Eckart}, F.~{Eisenhauer},
  M.~{Esselborn}, P.~{F{\'e}dou}, G.~{Finger}, P.~{Garcia}, C.~E. {Garcia
  Dabo}, R.~{Garcia Lopez}, E.~{Gendron}, R.~{Genzel}, S.~{Gillessen},
  F.~{Gonte}, P.~{Gordo}, M.~{Grould}, U.~{Gr{\"o}zinger}, S.~{Guieu},
  P.~{Haguenauer}, O.~{Hans}, X.~{Haubois}, M.~{Haug}, F.~{Haussmann},
  T.~{Henning}, S.~{Hippler}, M.~{Horrobin}, A.~{Huber}, Z.~{Hubert},
  N.~{Hubin}, C.~A. {Hummel}, G.~{Jakob}, A.~{Janssen}, L.~{Jochum},
  L.~{Jocou}, A.~{Kaufer}, S.~{Kellner}, S.~{Kendrew}, L.~{Kern},
  P.~{Kervella}, M.~{Kiekebusch}, R.~{Klein}, Y.~{Kok}, J.~{Kolb}, M.~{Kulas},
  S.~{Lacour}, V.~{Lapeyr{\`e}re}, B.~{Lazareff}, J.-B. {Le Bouquin},
  P.~{L{\`e}na}, R.~{Lenzen}, S.~{L{\'e}v{\^e}que}, M.~{Lippa}, Y.~{Magnard},
  L.~{Mehrgan}, M.~{Mellein}, A.~{M{\'e}rand}, J.~{Moreno-Ventas}, T.~{Moulin},
  E.~{M{\"u}ller}, F.~{M{\"u}ller}, U.~{Neumann}, S.~{Oberti}, T.~{Ott},
  L.~{Pallanca}, J.~{Panduro}, L.~{Pasquini}, T.~{Paumard}, I.~{Percheron},
  K.~{Perraut}, G.~{Perrin}, A.~{Pfl{\"u}ger}, O.~{Pfuhl}, T.~{Phan Duc}, P.~M.
  {Plewa}, D.~{Popovic}, S.~{Rabien}, A.~{Ram{\'{\i}}rez}, J.~{Ramos},
  C.~{Rau}, M.~{Riquelme}, R.-R. {Rohloff}, G.~{Rousset},
  J.~{Sanchez-Bermudez}, S.~{Scheithauer}, M.~{Sch{\"o}ller}, N.~{Schuhler},
  J.~{Spyromilio}, C.~{Straubmeier}, E.~{Sturm}, M.~{Suarez}, K.~R.~W.
  {Tristram}, N.~{Ventura}, F.~{Vincent}, I.~{Waisberg}, I.~{Wank}, J.~{Weber},
  E.~{Wieprecht}, M.~{Wiest}, E.~{Wiezorrek}, M.~{Wittkowski}, J.~{Woillez},
  B.~{Wolff}, S.~{Yazici}, D.~{Ziegler}, and G.~{Zins}, {First light for
  GRAVITY: Phase referencing optical interferometry for the Very Large
  Telescope Interferometer}, \emph{\aap}. 602:\penalty0 A94 (June, 2017).
\newblock \doi{10.1051/0004-6361/201730838}.

\bibitem{Tokunaga&Vacca2005}
A.~T. {Tokunaga} and W.~D. {Vacca}, {The Mauna Kea Observatories Near-Infrared
  Filter Set. III. Isophotal Wavelengths and Absolute Calibration},
  \emph{\pasp}. {\bf 117}, \penalty0 421--426 (Apr., 2005).
\newblock \doi{10.1086/429382}.

\bibitem{Lindegren2010}
L.~{Lindegren}, {High-accuracy positioning: astrometry}, \emph{ISSI Scientific
  Reports Series}. {\bf 9}, \penalty0 279--291,  (2010).

\bibitem{Sarazin&Tokovinin2002}
M.~{Sarazin} and A.~{Tokovinin}.
\newblock {The Statistics of Isoplanatic Angle and Adaptive Optics Time
  Constant derived from DIMM Data}.
\newblock In eds. E.~{Vernet}, R.~{Ragazzoni}, S.~{Esposito}, and N.~{Hubin},
  \emph{European Southern Observatory Conference and Workshop Proceedings},
  vol.~58, \emph{European Southern Observatory Conference and Workshop
  Proceedings}, p. 321,  (2002).

\bibitem{tenBrummelaar1996}
T.~A. {ten Brummelaar}, {Modeling atmospheric wave aberrations and astronomical
  instrumentation using the polynomials of Zernike}, \emph{Optics
  Communications}. {\bf 132}, \penalty0 329--342 (Feb., 1996).
\newblock \doi{10.1016/0030-4018(96)00407-5}.

\bibitem{Glindemann1997}
A.~{Glindemann}, {Relevant Parameters for Tip-Tilt Systems of Large
  Telescopes}, \emph{\pasp}. {\bf 109}, \penalty0 682--687 (June, 1997).
\newblock \doi{10.1086/133932}.

\bibitem{Glindemann2011}
A.~{Glindemann}, \emph{{Principles of Stellar Interferometry}}. 2011.
\newblock \doi{10.1007/978-3-642-15028-9}.

\bibitem{Benson+2010}
J.~A. {Benson}, D.~J. {Hutter}, R.~T. {Zavala}, H.~C. {Harris}, P.~D.
  {Shankland}, and K.~J. {Johnston}.
\newblock {From fringes to the USNO Navy Prototype Optical Interferometer
  Astrometric Catalog}.
\newblock In \emph{Optical and Infrared Interferometry II}, vol. 7734,
  \emph{\procspie}, p. 77343K (July, 2010).
\newblock \doi{10.1117/12.858244}.

\bibitem{Shao&Colavita1992}
M.~{Shao} and M.~M. {Colavita}, {Potential of long-baseline infrared
  interferometry for narrow-angle astrometry}, \emph{\aap}. {\bf 262},
  \penalty0 353--358 (Aug., 1992).

\bibitem{Lindegren1980}
L.~{Lindegren}, {Atmospheric limitations of narrow-field optical astrometry},
  \emph{\aap}. {\bf 89}, \penalty0 41--47 (Sept., 1980).

\bibitem{Conan+1995}
J.-M. {Conan}, G.~{Rousset}, and P.-Y. {Madec}, {Wave-front temporal spectra in
  high-resolution imaging through turbulence.}, \emph{Journal of the Optical
  Society of America A}. {\bf 12}, \penalty0 1559--1570 (July, 1995).
\newblock \doi{10.1364/JOSAA.12.001559}.

\bibitem{Lazorenko&Lazorenko2004}
P.~F. {Lazorenko} and G.~A. {Lazorenko}, {Filtration of atmospheric noise in
  narrow-field astrometry with very large telescopes}, \emph{\aap}. {\bf 427},
  \penalty0 1127--1143 (Dec., 2004).
\newblock \doi{10.1051/0004-6361:20041481}.

\bibitem{Cameron+2009}
P.~B. {Cameron}, M.~C. {Britton}, and S.~R. {Kulkarni}, {Precision Astrometry
  With Adaptive Optics}, \emph{\aj}. {\bf 137}, \penalty0 83--93 (Jan., 2009).
\newblock \doi{10.1088/0004-6256/137/1/83}.

\bibitem{Colavita2009}
M.~M. {Colavita}, {Adverse effects in dual-feed interferometry}, \emph{\nar}.
  {\bf 53}, \penalty0 344--352 (Nov., 2009).
\newblock \doi{10.1016/j.newar.2010.07.004}.

\bibitem{Buscher+2009}
D.~{Buscher}, F.~{Baron}, and C.~{Haniff}, {Minimizing the Effects of
  Polarization Crosstalk on the Imaging Fidelity of an Optical Interferometer},
  \emph{\pasp}. {\bf 121}, \penalty0 45 (Jan., 2009).
\newblock \doi{10.1086/597127}.

\bibitem{Colavita+2013}
M.~M. {Colavita}, P.~L. {Wizinowich}, R.~L. {Akeson}, S.~{Ragland}, J.~M.
  {Woillez}, R.~{Millan-Gabet}, E.~{Serabyn}, M.~{Abajian}, D.~S. {Acton},
  E.~{Appleby}, J.~W. {Beletic}, C.~A. {Beichman}, J.~{Bell}, B.~C. {Berkey},
  J.~{Berlin}, A.~F. {Boden}, A.~J. {Booth}, R.~{Boutell}, F.~H. {Chaffee},
  D.~{Chan}, J.~{Chin}, J.~{Chock}, R.~{Cohen}, A.~{Cooper}, S.~L. {Crawford},
  M.~J. {Creech-Eakman}, W.~{Dahl}, G.~{Eychaner}, J.~L. {Fanson},
  C.~{Felizardo}, J.~I. {Garcia-Gathright}, J.~T. {Gathright}, G.~{Hardy},
  H.~{Henderson}, J.~S. {Herstein}, M.~{Hess}, E.~E. {Hovland}, M.~A.
  {Hrynevych}, E.~{Johansson}, R.~L. {Johnson}, J.~{Kelley}, R.~{Kendrick},
  C.~D. {Koresko}, P.~{Kurpis}, D.~{Le Mignant}, H.~A. {Lewis}, E.~R. {Ligon},
  W.~{Lupton}, D.~{McBride}, D.~W. {Medeiros}, B.~P. {Mennesson}, J.~D.
  {Moore}, D.~{Morrison}, C.~{Nance}, C.~{Neyman}, A.~{Niessner}, C.~G.
  {Paine}, D.~L. {Palmer}, T.~{Panteleeva}, M.~{Papin}, B.~{Parvin},
  L.~{Reder}, A.~{Rudeen}, T.~{Saloga}, A.~{Sargent}, M.~{Shao}, B.~{Smith},
  R.~F. {Smythe}, P.~{Stomski}, K.~R. {Summers}, M.~R. {Swain}, P.~{Swanson},
  R.~{Thompson}, K.~{Tsubota}, A.~{Tumminello}, C.~{Tyau}, G.~T. {van Belle},
  G.~{Vasisht}, J.~{Vause}, F.~{Vescelus}, J.~{Walker}, J.~K. {Wallace},
  U.~{Wehmeier}, and E.~{Wetherell}, {The Keck Interferometer}, \emph{\pasp}.
  {\bf 125}, \penalty0 1226 (Oct., 2013).
\newblock \doi{10.1086/673475}.

\bibitem{Woillez+2014b}
J.~{Woillez}, R.~{Abuter}, L.~{Andolfato}, J.-P. {Berger}, H.~{Bonnet},
  F.~{Delplancke}, F.~{Derie}, N.~{Di Lieto}, S.~{Guniat}, A.~{M{\'e}rand},
  T.~P. {Duc}, C.~{Schmid}, N.~{Schuhler}, T.~{Henning}, R.~{Launhardt},
  F.~{Pepe}, D.~{Queloz}, A.~{Quirrenbach}, S.~{Reffert}, J.~{Sahlmann}, and
  D.~{Segransan}.
\newblock {Improving the astrometric performance of VLTI-PRIMA}.
\newblock In \emph{Optical and Infrared Interferometry IV}, vol. 9146,
  \emph{\procspie}, p. 91461H (July, 2014).
\newblock \doi{10.1117/12.2054723}.

\bibitem{Lazareff+2014}
B.~{Lazareff}, N.~{Blind}, L.~{Jocou}, F.~{Eisenhauer}, K.~{Perraut},
  S.~{Lacour}, F.~{Delplancke}, M.~{Schoeller}, A.~{Amorim}, W.~{Brandner},
  G.~{Perrin}, and C.~{Straubmeier}.
\newblock {Telescope birefringence and phase errors in the Gravity instrument
  at the VLT interferometer}.
\newblock In \emph{Optical and Infrared Interferometry IV}, vol. 9146,
  \emph{\procspie}, p. 91460X (July, 2014).
\newblock \doi{10.1117/12.2056304}.

\bibitem{Sahlmann+2013}
J.~{Sahlmann}, T.~{Henning}, D.~{Queloz}, A.~{Quirrenbach}, N.~M. {Elias},
  R.~{Launhardt}, F.~{Pepe}, S.~{Reffert}, D.~{S{\'e}gransan}, J.~{Setiawan},
  R.~{Abuter}, L.~{Andolfato}, P.~{Bizenberger}, H.~{Baumeister},
  B.~{Chazelas}, F.~{Delplancke}, F.~{D{\'e}rie}, N.~{Di Lieto}, T.~P. {Duc},
  M.~{Fleury}, U.~{Graser}, A.~{Kaminski}, R.~{K{\"o}hler},
  S.~{L{\'e}v{\^e}que}, C.~{Maire}, D.~{M{\'e}gevand}, A.~{M{\'e}rand},
  Y.~{Michellod}, J.-M. {Moresmau}, M.~{Mohler}, A.~{M{\"u}ller},
  P.~{M{\"u}llhaupt}, V.~{Naranjo}, L.~{Sache}, Y.~{Salvade}, C.~{Schmid},
  N.~{Schuhler}, T.~{Schulze-Hartung}, D.~{Sosnowska}, B.~{Tubbs}, G.~T. {van
  Belle}, K.~{Wagner}, L.~{Weber}, L.~{Zago}, and N.~{Zimmerman}, {The ESPRI
  project: astrometric exoplanet search with PRIMA. I. Instrument description
  and performance of first light observations}, \emph{\aap}. 551:\penalty0 A52
  (Mar., 2013).
\newblock \doi{10.1051/0004-6361/201220569}.

\bibitem{Woillez&Lacour2013}
J.~{Woillez} and S.~{Lacour}, {Wide-angle, Narrow-angle, and Imaging Baselines
  of Optical Long-baseline Interferometers}, \emph{\apj}. 764:\penalty0 109
  (Feb., 2013).
\newblock \doi{10.1088/0004-637X/764/1/109}.

\bibitem{GaiaCollaboration+2016}
{Gaia Collaboration}, T.~{Prusti}, J.~H.~J. {de Bruijne}, A.~G.~A. {Brown},
  A.~{Vallenari}, C.~{Babusiaux}, C.~A.~L. {Bailer-Jones}, U.~{Bastian},
  M.~{Biermann}, D.~W. {Evans}, and et~al., {The Gaia mission}, \emph{\aap}.
  595:\penalty0 A1 (Nov., 2016).
\newblock \doi{10.1051/0004-6361/201629272}.

\bibitem{Hrynevych+2004}
M.~A. {Hrynevych}, E.~R. {Ligon}, III, and M.~M. {Colavita}.
\newblock {Baseline monitoring for astrometric interferometry}.
\newblock In ed. W.~A. {Traub}, \emph{New Frontiers in Stellar Interferometry},
  vol. 5491, \emph{\procspie}, p. 1649 (Oct., 2004).
\newblock \doi{10.1117/12.552320}.

\bibitem{Woillez+2010}
J.~{Woillez}, R.~{Akeson}, M.~{Colavita}, J.~{Eisner}, A.~{Ghez}, J.~{Graham},
  L.~{Hillenbrand}, R.~{Millan-Gabet}, J.~{Monnier}, J.-U. {Pott},
  S.~{Ragland}, P.~{Wizinowich}, E.~{Appleby}, B.~{Berkey}, A.~{Cooper},
  C.~{Felizardo}, J.~{Herstein}, M.~{Hrynevych}, O.~{Martin}, D.~{Medeiros},
  D.~{Morrison}, T.~{Panteleeva}, B.~{Smith}, K.~{Summers}, K.~{Tsubota},
  C.~{Tyau}, and E.~{Wetherell}.
\newblock {ASTRA: astrometry and phase-referencing astronomy on the Keck
  interferometer}.
\newblock In \emph{Optical and Infrared Interferometry II}, vol. 7734,
  \emph{\procspie}, p. 773412 (July, 2010).
\newblock \doi{10.1117/12.857740}.

\bibitem{Shao+1988}
M.~{Shao}, M.~M. {Colavita}, B.~E. {Hines}, D.~H. {Staelin}, D.~J. {Hutter},
  K.~J. {Johnston}, D.~{Mozurkewich}, R.~S. {Simon}, J.~L. {Hershey}, J.~A.
  {Hughes}, and G.~H. {Kaplan}, {The Mark III stellar interferometer},
  \emph{\aap}. {\bf 193}, \penalty0 357--371 (Mar., 1988).

\bibitem{Colavita1994}
M.~M. {Colavita}, {Measurement of the Atmospheric Limit to Narrow Angle
  Interferometric Astrometry Using the Mark-Iii Stellar Interferometer},
  \emph{\aap}. {\bf 283}, \penalty0 1027 (Mar., 1994).

\bibitem{Colavita+1999}
M.~M. {Colavita}, J.~K. {Wallace}, B.~E. {Hines}, Y.~{Gursel}, F.~{Malbet},
  D.~L. {Palmer}, X.~P. {Pan}, M.~{Shao}, J.~W. {Yu}, A.~F. {Boden}, P.~J.
  {Dumont}, J.~{Gubler}, C.~D. {Koresko}, S.~R. {Kulkarni}, B.~F. {Lane}, D.~W.
  {Mobley}, and G.~T. {van Belle}, {The Palomar Testbed Interferometer},
  \emph{\apj}. {\bf 510}, \penalty0 505--521 (Jan., 1999).
\newblock \doi{10.1086/306579}.

\bibitem{Shao+1999}
M.~{Shao}, A.~F. {Boden}, M.~M. {Colavita}, B.~F. {Lane}, P.~R. {Lawson}, and
  {PTI Collaboration}.
\newblock {Differential Astrometry of the 61 Cygni System with the Palomar
  Testbed Interferometer}.
\newblock In \emph{American Astronomical Society Meeting Abstracts}, vol.~31,
  \emph{Bulletin of the American Astronomical Society}, p. 1504 (Dec., 1999).

\bibitem{Lane+2000}
B.~F. {Lane}, M.~M. {Colavita}, A.~F. {Boden}, and P.~R. {Lawson}.
\newblock {Palomar Testbed Interferometer: update}.
\newblock In eds. P.~{L{\'e}na} and A.~{Quirrenbach}, \emph{Interferometry in
  Optical Astronomy}, vol. 4006, \emph{\procspie}, pp. 452--458 (July, 2000).
\newblock \doi{10.1117/12.390239}.

\bibitem{Muterspaugh+2010}
M.~W. {Muterspaugh}, B.~F. {Lane}, S.~R. {Kulkarni}, M.~{Konacki}, B.~F.
  {Burke}, M.~M. {Colavita}, M.~{Shao}, S.~J. {Wiktorowicz}, and
  J.~{O'Connell}, {The Phases Differential Astrometry Data Archive. I.
  Measurements and Description}, \emph{\aj}. {\bf 140}, \penalty0 1579--1622
  (Dec., 2010).
\newblock \doi{10.1088/0004-6256/140/6/1579}.

\bibitem{Lachaume&Berger2013}
R.~{Lachaume} and J.-P. {Berger}, {Bandwidth smearing in infrared long-baseline
  interferometry. Application to stellar companion search in fringe-scanning
  mode}, \emph{\mnras}. {\bf 435}, \penalty0 2501--2519 (Nov., 2013).
\newblock \doi{10.1093/mnras/stt1462}.

\bibitem{Colavita+1998}
M.~M. {Colavita}, A.~F. {Boden}, S.~L. {Crawford}, A.~B. {Meinel}, M.~{Shao},
  P.~N. {Swanson}, G.~T. {van Belle}, G.~{Vasisht}, J.~M. {Walker}, J.~K.
  {Wallace}, and P.~L. {Wizinowich}.
\newblock {Keck Interferometer}.
\newblock In ed. R.~D. {Reasenberg}, \emph{Astronomical Interferometry}, vol.
  3350, \emph{\procspie}, pp. 776--784 (July, 1998).
\newblock \doi{10.1117/12.317145}.

\bibitem{Bell+2004}
J.~{Bell}, J.~M. {Walker}, P.~L. {Wizinowich}, K.~{Tsubota}, A.~C. {Rudeen},
  D.~{McBride}, K.~K. {Kinoshita}, M.~{Hrynevych}, P.~{Goude}, M.~M.
  {Colavita}, J.~H. {Kelley}, G.~T. {van Belle}, R.~{Brunswick}, J.~K.
  {Little}, and C.~H. {Smith}.
\newblock {Outrigger telescopes for narrow-angle astrometry}.
\newblock In ed. J.~M. {Oschmann}, Jr., \emph{Ground-based Telescopes}, vol.
  5489, \emph{\procspie}, pp. 962--973 (Oct., 2004).
\newblock \doi{10.1117/12.552403}.

\bibitem{Pott+2009}
J.-U. {Pott}, J.~{Woillez}, R.~L. {Akeson}, B.~{Berkey}, M.~M. {Colavita},
  A.~{Cooper}, J.~A. {Eisner}, A.~M. {Ghez}, J.~R. {Graham}, L.~{Hillenbrand},
  M.~{Hrynewych}, D.~{Medeiros}, R.~{Millan-Gabet}, J.~{Monnier},
  D.~{Morrison}, T.~{Panteleeva}, E.~{Quataert}, B.~{Randolph}, B.~{Smith},
  K.~{Summers}, K.~{Tsubota}, C.~{Tyau}, N.~{Weinberg}, E.~{Wetherell}, and
  P.~L. {Wizinowich}, {Astrometry with the Keck Interferometer: The ASTRA
  project and its science}, \emph{\nar}. {\bf 53}, \penalty0 363--372 (Nov.,
  2009).
\newblock \doi{10.1016/j.newar.2010.07.009}.

\bibitem{Woillez+2014a}
J.~{Woillez}, P.~{Wizinowich}, R.~{Akeson}, M.~{Colavita}, J.~{Eisner},
  R.~{Millan-Gabet}, J.~D. {Monnier}, J.-U. {Pott}, and S.~{Ragland}, {First
  Faint Dual-field Off-axis Observations in Optical Long Baseline
  Interferometry}, \emph{\apj}. 783:\penalty0 104 (Mar., 2014).
\newblock \doi{10.1088/0004-637X/783/2/104}.

\bibitem{Ragland+2012}
S.~{Ragland}, R.~{Akeson}, M.~{Colavita}, R.~{Millan-Gabet}, T.~{Pantaleeva},
  B.~{Smith}, K.~{Summers}, P.~{Wizinowich}, J.~{Woillez}, E.~{Appleby},
  A.~{Cooper}, C.~{Felizardo}, J.~{Herstein}, D.~{Morrison}, K.~{Tsubota}, and
  C.~{Tyau}.
\newblock {Recent progress at the Keck interferometer}.
\newblock In \emph{Optical and Infrared Interferometry III}, vol. 8445,
  \emph{\procspie}, p. 84450C (July, 2012).
\newblock \doi{10.1117/12.925462}.

\bibitem{Kok13}
Y.~{Kok}, M.~J. {Ireland}, P.~G. {Tuthill}, J.~G. {Robertson}, B.~A.
  {Warrington}, A.~C. {Rizzuto}, and W.~J. {Tango}, {Phase-Referenced
  Interferometry and Narrow-Angle Astrometry with SUSI}, \emph{Journal of
  Astronomical Instrumentation}. 2:\penalty0 1340011,  (2013).
\newblock \doi{10.1142/S2251171713400114}.

\bibitem{Delplancke2008}
F.~{Delplancke}, {The PRIMA facility phase-referenced imaging and
  micro-arcsecond astrometry}, \emph{\nar}. {\bf 52}, \penalty0 199--207 (June,
  2008).
\newblock \doi{10.1016/j.newar.2008.04.016}.

\bibitem{Delplancke+2004}
F.~{Delplancke}, J.~{Nijenhuis}, H.~{de Man}, L.~{Andolfato}, R.~{Treichel},
  J.~{Hopman}, and F.~{Derie}.
\newblock {Star separator system for the dual-field capability (PRIMA) of the
  VLTI}.
\newblock In ed. W.~A. {Traub}, \emph{New Frontiers in Stellar Interferometry},
  vol. 5491, \emph{\procspie}, p. 1528 (Oct., 2004).
\newblock \doi{10.1117/12.551873}.

\bibitem{Pepe+2008}
F.~{Pepe}, D.~{Queloz}, T.~{Henning}, A.~{Quirrenbach}, F.~{Delplancke},
  L.~{Andolfato}, H.~{Baumeister}, P.~{Bizenberger}, H.~{Bleuler},
  B.~{Chazelas}, F.~{D{\'e}rie}, L.~{Di Lieto}, T.~P. {Duc}, O.~{Duvanel},
  M.~{Fleury}, D.~{Gillet}, U.~{Graser}, F.~{Koch}, R.~{Launhardt}, C.~{Maire},
  D.~{M{\'e}gevand}, Y.~{Michellod}, J.-M. {Moresmau}, P.~{M{\"u}llhaupt},
  V.~{Naranjo}, L.~{Sache}, Y.~{Salvad{\'e}}, G.~{Simond}, D.~{Sosnowska},
  K.~{Wagner}, and L.~{Zago}.
\newblock {The ESPRI Project: differential delay lines for PRIMA}.
\newblock In \emph{Optical and Infrared Interferometry}, vol. 7013,
  \emph{\procspie}, p. 70130P (July, 2008).
\newblock \doi{10.1117/12.788169}.

\bibitem{Sahlmann+2009}
J.~{Sahlmann}, S.~{M{\'e}nardi}, R.~{Abuter}, M.~{Accardo}, S.~{Mottini}, and
  F.~{Delplancke}, {The PRIMA fringe sensor unit}, \emph{\aap}. {\bf 507},
  \penalty0 1739--1757 (Dec., 2009).
\newblock \doi{10.1051/0004-6361/200912271}.

\bibitem{Leveque+2003}
S.~A. {Leveque}, R.~{Wilhelm}, Y.~{Salvade}, O.~{Scherler}, and
  R.~{Daendliker}.
\newblock {Toward nanometer accuracy laser metrology for phase-referenced
  interferometry with the VLTI}.
\newblock In ed. W.~A. {Traub}, \emph{Interferometry for Optical Astronomy II},
  vol. 4838, \emph{\procspie}, pp. 983--994 (Feb., 2003).
\newblock \doi{10.1117/12.457173}.

\bibitem{Launhardt+2008}
R.~{Launhardt}, D.~{Queloz}, T.~{Henning}, A.~{Quirrenbach}, F.~{Delplancke},
  L.~{Andolfato}, H.~{Baumeister}, P.~{Bizenberger}, H.~{Bleuler},
  B.~{Chazelas}, F.~{D{\'e}rie}, L.~{Di Lieto}, T.~P. {Duc}, O.~{Duvanel},
  N.~M. {Elias}, II, M.~{Fluery}, R.~{Geisler}, D.~{Gillet}, U.~{Graser},
  F.~{Koch}, R.~{K{\"o}hler}, C.~{Maire}, D.~{M{\'e}gevand}, Y.~{Michellod},
  J.-M. {Moresmau}, A.~{M{\"u}ller}, P.~{M{\"u}llhaupt}, V.~{Naranjo},
  F.~{Pepe}, S.~{Reffert}, L.~{Sache}, D.~{S{\'e}gransan}, Y.~{Salvad{\'e}},
  T.~{Schulze-Hartung}, J.~{Setiawan}, G.~{Simond}, D.~{Sosnowska}, I.~{Stilz},
  B.~{Tubbs}, K.~{Wagner}, L.~{Weber}, P.~{Weise}, and L.~{Zago}.
\newblock {The ESPRI project: astrometric exoplanet search with PRIMA}.
\newblock In \emph{Optical and Infrared Interferometry}, vol. 7013,
  \emph{\procspie}, p. 70132I (July, 2008).
\newblock \doi{10.1117/12.789318}.

\bibitem{Eisenhauer+2011}
F.~{Eisenhauer}, G.~{Perrin}, W.~{Brandner}, C.~{Straubmeier}, K.~{Perraut},
  A.~{Amorim}, M.~{Sch{\"o}ller}, S.~{Gillessen}, P.~{Kervella}, M.~{Benisty},
  C.~{Araujo-Hauck}, L.~{Jocou}, J.~{Lima}, G.~{Jakob}, M.~{Haug},
  Y.~{Cl{\'e}net}, T.~{Henning}, A.~{Eckart}, J.-P. {Berger}, P.~{Garcia},
  R.~{Abuter}, S.~{Kellner}, T.~{Paumard}, S.~{Hippler}, S.~{Fischer},
  T.~{Moulin}, J.~{Villate}, G.~{Avila}, A.~{Gr{\"a}ter}, S.~{Lacour},
  A.~{Huber}, M.~{Wiest}, A.~{Nolot}, P.~{Carvas}, R.~{Dorn}, O.~{Pfuhl},
  E.~{Gendron}, S.~{Kendrew}, S.~{Yazici}, S.~{Anton}, Y.~{Jung}, M.~{Thiel},
  {\'E}.~{Choquet}, R.~{Klein}, P.~{Teixeira}, P.~{Gitton}, D.~{Moch},
  F.~{Vincent}, N.~{Kudryavtseva}, S.~{Str{\"o}bele}, S.~{Sturm},
  P.~{F{\'e}dou}, R.~{Lenzen}, P.~{Jolley}, C.~{Kister}, V.~{Lapeyr{\`e}re},
  V.~{Naranjo}, C.~{Lucuix}, R.~{Hofmann}, F.~{Chapron}, U.~{Neumann},
  L.~{Mehrgan}, O.~{Hans}, G.~{Rousset}, J.~{Ramos}, M.~{Suarez}, R.~{Lederer},
  J.-M. {Reess}, R.-R. {Rohloff}, P.~{Haguenauer}, H.~{Bartko}, A.~{Sevin},
  K.~{Wagner}, J.-L. {Lizon}, S.~{Rabien}, C.~{Collin}, G.~{Finger},
  R.~{Davies}, D.~{Rouan}, M.~{Wittkowski}, K.~{Dodds-Eden}, D.~{Ziegler},
  F.~{Cassaing}, H.~{Bonnet}, M.~{Casali}, R.~{Genzel}, and P.~{Lena},
  {GRAVITY: Observing the Universe in Motion}, \emph{The Messenger}. {\bf 143},
  \penalty0 16--24 (Mar., 2011).

\bibitem{Gillessen+2010}
S.~{Gillessen}, F.~{Eisenhauer}, G.~{Perrin}, W.~{Brandner}, C.~{Straubmeier},
  K.~{Perraut}, A.~{Amorim}, M.~{Sch{\"o}ller}, C.~{Araujo-Hauck}, H.~{Bartko},
  H.~{Baumeister}, J.-P. {Berger}, P.~{Carvas}, F.~{Cassaing}, F.~{Chapron},
  E.~{Choquet}, Y.~{Clenet}, C.~{Collin}, A.~{Eckart}, P.~{Fedou},
  S.~{Fischer}, E.~{Gendron}, R.~{Genzel}, P.~{Gitton}, F.~{Gonte},
  A.~{Gr{\"a}ter}, P.~{Haguenauer}, M.~{Haug}, X.~{Haubois}, T.~{Henning},
  S.~{Hippler}, R.~{Hofmann}, L.~{Jocou}, S.~{Kellner}, P.~{Kervella},
  R.~{Klein}, N.~{Kudryavtseva}, S.~{Lacour}, V.~{Lapeyrere}, W.~{Laun},
  P.~{Lena}, R.~{Lenzen}, J.~{Lima}, D.~{Moratschke}, D.~{Moch}, T.~{Moulin},
  V.~{Naranjo}, U.~{Neumann}, A.~{Nolot}, T.~{Paumard}, O.~{Pfuhl},
  S.~{Rabien}, J.~{Ramos}, J.~M. {Rees}, R.-R. {Rohloff}, D.~{Rouan},
  G.~{Rousset}, A.~{Sevin}, M.~{Thiel}, K.~{Wagner}, M.~{Wiest}, S.~{Yazici},
  and D.~{Ziegler}.
\newblock {GRAVITY: a four-telescope beam combiner instrument for the VLTI}.
\newblock In \emph{Optical and Infrared Interferometry II}, vol. 7734,
  \emph{\procspie}, p. 77340Y (July, 2010).
\newblock \doi{10.1117/12.856689}.

\bibitem{Gillessen+2012}
S.~{Gillessen}, M.~{Lippa}, F.~{Eisenhauer}, O.~{Pfuhl}, M.~{Haug},
  S.~{Kellner}, T.~{Ott}, E.~{Wieprecht}, E.~{Sturm}, F.~{Hau{\ss}mann}, C.~F.
  {Kister}, D.~{Moch}, and M.~{Thiel}.
\newblock {GRAVITY: metrology}.
\newblock In \emph{Optical and Infrared Interferometry III}, vol. 8445,
  \emph{\procspie}, p. 84451O (July, 2012).
\newblock \doi{10.1117/12.926813}.

\bibitem{Jocou+2014}
L.~{Jocou}, K.~{Perraut}, T.~{Moulin}, Y.~{Magnard}, P.~{Labeye}, V.~{Lapras},
  A.~{Nolot}, G.~{Perrin}, F.~{Eisenhauer}, C.~{Holmes}, A.~{Amorim},
  W.~{Brandner}, and C.~{Straubmeier}.
\newblock {The beam combiners of Gravity VLTI instrument: concept, development,
  and performance in laboratory}.
\newblock In \emph{Optical and Infrared Interferometry IV}, vol. 9146,
  \emph{\procspie}, p. 91461J (July, 2014).
\newblock \doi{10.1117/12.2054159}.

\bibitem{Anugu+2016}
N.~{Anugu}, P.~{Garcia}, A.~{Amorim}, E.~{Wiezorrek}, E.~{Wieprecht},
  F.~{Eisenhauer}, T.~{Ott}, O.~{Pfuhl}, P.~{Gordo}, G.~{Perrin},
  W.~{Brandner}, C.~{Straubmeier}, and K.~{Perraut}.
\newblock {GRAVITY acquisition camera: characterization results}.
\newblock In \emph{Optical and Infrared Interferometry and Imaging V}, vol.
  9907, \emph{\procspie}, p. 990727 (Aug., 2016).
\newblock \doi{10.1117/12.2233315}.

\bibitem{Lippa+2016}
M.~{Lippa}, S.~{Gillessen}, N.~{Blind}, Y.~{Kok}, {\c S}.~{Yaz{\i}c{\i}},
  J.~{Weber}, O.~{Pfuhl}, M.~{Haug}, S.~{Kellner}, E.~{Wieprecht},
  F.~{Eisenhauer}, R.~{Genzel}, O.~{Hans}, F.~{Hau{\ss}mann}, D.~{Huber},
  T.~{Kratschmann}, T.~{Ott}, M.~{Plattner}, C.~{Rau}, E.~{Sturm},
  I.~{Waisberg}, E.~{Wiezorrek}, G.~{Perrin}, K.~{Perraut}, W.~{Brandner},
  C.~{Straubmeier}, and A.~{Amorim}.
\newblock {The metrology system of the VLTI instrument GRAVITY}.
\newblock In \emph{Optical and Infrared Interferometry and Imaging V}, vol.
  9907, \emph{\procspie}, p. 990722 (Aug., 2016).
\newblock \doi{10.1117/12.2232272}.

\bibitem{Lacour+2014}
S.~{Lacour}, F.~{Eisenhauer}, S.~{Gillessen}, O.~{Pfuhl}, J.~{Woillez},
  H.~{Bonnet}, G.~{Perrin}, B.~{Lazareff}, S.~{Rabien}, V.~{Lapeyr{\`e}re},
  Y.~{Cl{\'e}net}, P.~{Kervella}, and Y.~{Kok}, {Reaching micro-arcsecond
  astrometry with long baseline optical interferometry. Application to the
  GRAVITY instrument}, \emph{\aap}. 567:\penalty0 A75 (July, 2014).
\newblock \doi{10.1051/0004-6361/201423940}.

\bibitem{Martina18}
{GRAVITY collaboration}, M.~{Karl}, O.~{Pfuhl}, F.~{Eisenhauer}, R.~{Genzel},
  R.~{Grellmann}, M.~{Habibi}, R.~{Abuter}, M.~{Accardo}, A.~{Amorim},
  N.~{Anugu}, G.~{{\'A}vila}, M.~{Benisty}, J.-P. {Berger}, N.~{Bland},
  H.~{Bonnet}, P.~{Bourget}, W.~{Brandner}, R.~{Brast}, A.~{Buron}, A.~C.~o.
  {Garatti}, F.~{Chapron}, Y.~{Cl{\'e}net}, C.~{Collin}, V.~{Coud{\'e} du
  Foresto}, W.-J. {de Wit}, T.~{de Zeeuw}, C.~{Deen},
  F.~{Delplancke-Str{\"o}bele}, R.~{Dembet}, F.~{Derie}, J.~{Dexter},
  G.~{Duvert}, M.~{Ebert}, A.~{Eckart}, M.~{Esselborn}, P.~{F{\'e}dou},
  G.~{Finger}, P.~{Garcia}, C.~E. {Garcia Dabo}, R.~{Garcia Lopez}, F.~{Gao},
  {\'E}.~{Gandron}, S.~{Gillessen}, F.~{Gont{\'e}}, P.~{Gordo},
  U.~{Gr{\"o}zinger}, P.~{Guajardo}, S.~{Guieu}, P.~{Haguenauer}, O.~{Hans},
  X.~{Haubois}, M.~{Haug}, F.~{Hau{\ss}mann}, T.~{Henning}, S.~{Hippler},
  M.~{Horrobin}, A.~{Huber}, Z.~{Hubert}, N.~{Hubin}, C.~A. {Hummel},
  G.~{Jakob}, L.~{Jochum}, L.~{Jocou}, A.~{Kaufer}, S.~{Kellner}, S.~{Kandrew},
  L.~{Kern}, P.~{Kervella}, M.~{Kiekebusch}, R.~{Klein}, R.~{K{\"o}hler},
  J.~{Kolb}, M.~{Kulas}, S.~{Lacour}, V.~{Lapeyr{\`e}re}, B.~{Lazareff}, J.-B.
  {Le Bouquin}, P.~{L{\'e}na}, R.~{Lenzen}, S.~{L{\'e}v{\^e}que}, C.-C. {Lin},
  M.~{Lippa}, Y.~{Magnard}, L.~{Mehrgan}, A.~{M{\'e}rand}, T.~{Moulin},
  E.~{M{\"u}ller}, F.~{M{\"u}ller}, U.~{Neumann}, S.~{Oberti}, T.~{Ott},
  L.~{Pallanca}, J.~{Panduro}, L.~{Pasquini}, T.~{Paumard}, I.~{Percheron},
  K.~{Perraut}, G.~{Perrin}, A.~{Pfl{\"u}ger}, T.~{Phan Duc}, P.~M. {Plewa},
  D.~{Popovic}, S.~{Rabien}, A.~{Ram{\'\i}rez}, J.~{Ramos}, C.~{Rau},
  M.~{Riquelme}, G.~{Rodr{\'\i}guez-Coira}, R.-R. {Rohloff}, A.~{Rosales},
  G.~{Rousset}, J.~{Sanchez-Bermudez}, S.~{Scheithauer}, M.~{Sch{\"o}ller},
  N.~{Schuhler}, J.~{Spyromilio}, O.~{Straub}, C.~{Straubmeier}, E.~{Sturm},
  M.~{Suarez}, K.~R.~W. {Tristram}, N.~{Ventura}, F.~{Vincent}, I.~{Waisberg},
  I.~{Wank}, F.~{Widmann}, E.~{Wieprecht}, M.~{Wiest}, E.~{Wiezorrek},
  M.~{Wittkowski}, J.~{Woillez}, B.~{Wolff}, S.~{Yazici}, D.~{Ziegler}, and
  G.~{Zins}, {Multiple Star Systems in the Orion Nebula}, \emph{ArXiv
  e-prints}. art. arXiv:1809.10376 (Sept., 2018).

\bibitem{Widmann18}
F.~{Widmann}, F.~{Eisenhauer}, G.~{Perrin}, W.~{Brandner}, C.~{Straubmeier},
  K.~{Perraut}, A.~{Amorim}, M.~{Sch{\"o}ller}, F.~{Gao}, R.~{Genzel},
  S.~{Gillessen}, M.~{Karl}, S.~{Lacour}, M.~{Lippa}, T.~{Ott}, O.~{Pfuhl},
  P.~{Plewa}, and I.~{Waisberg}.
\newblock {Improving GRAVITY towards observations of faint targets}.
\newblock In \emph{Society of Photo-Optical Instrumentation Engineers (SPIE)
  Conference Series}, vol. 10701, p. 107010K (July, 2018).
\newblock \doi{10.1117/12.2312118}.

\bibitem{Gravity18}
{Gravity Collaboration}, R.~{Abuter}, A.~{Amorim}, N.~{Anugu},
  M.~{Baub{\"o}ck}, M.~{Benisty}, J.~P. {Berger}, N.~{Blind}, H.~{Bonnet},
  W.~{Brandner}, A.~{Buron}, C.~{Collin}, F.~{Chapron}, Y.~{Cl{\'e}net},
  V.~{Coud{\'e} Du Foresto}, P.~T. {de Zeeuw}, C.~{Deen},
  F.~{Delplancke-Str{\"o}bele}, R.~{Dembet}, J.~{Dexter}, G.~{Duvert},
  A.~{Eckart}, F.~{Eisenhauer}, G.~{Finger}, N.~M. {F{\"o}rster Schreiber},
  P.~{F{\'e}dou}, P.~{Garcia}, R.~{Garcia Lopez}, F.~{Gao}, E.~{Gendron},
  R.~{Genzel}, S.~{Gillessen}, P.~{Gordo}, M.~{Habibi}, X.~{Haubois},
  M.~{Haug}, F.~{Hau{\ss}mann}, T.~{Henning}, S.~{Hippler}, M.~{Horrobin},
  Z.~{Hubert}, N.~{Hubin}, A.~{Jimenez Rosales}, L.~{Jochum}, K.~{Jocou},
  A.~{Kaufer}, S.~{Kellner}, S.~{Kendrew}, P.~{Kervella}, Y.~{Kok}, M.~{Kulas},
  S.~{Lacour}, V.~{Lapeyr{\`e}re}, B.~{Lazareff}, J.-B. {Le Bouquin},
  P.~{L{\'e}na}, M.~{Lippa}, R.~{Lenzen}, A.~{M{\'e}rand}, E.~{M{\"u}ler},
  U.~{Neumann}, T.~{Ott}, L.~{Palanca}, T.~{Paumard}, L.~{Pasquini},
  K.~{Perraut}, G.~{Perrin}, O.~{Pfuhl}, P.~M. {Plewa}, S.~{Rabien},
  A.~{Ram{\'{\i}}rez}, J.~{Ramos}, C.~{Rau}, G.~{Rodr{\'{\i}}guez-Coira}, R.-R.
  {Rohloff}, G.~{Rousset}, J.~{Sanchez-Bermudez}, S.~{Scheithauer},
  M.~{Sch{\"o}ller}, N.~{Schuler}, J.~{Spyromilio}, O.~{Straub},
  C.~{Straubmeier}, E.~{Sturm}, L.~J. {Tacconi}, K.~R.~W. {Tristram},
  F.~{Vincent}, S.~{von Fellenberg}, I.~{Wank}, I.~{Waisberg}, F.~{Widmann},
  E.~{Wieprecht}, M.~{Wiest}, E.~{Wiezorrek}, J.~{Woillez}, S.~{Yazici},
  D.~{Ziegler}, and G.~{Zins}, {Detection of the gravitational redshift in the
  orbit of the star S2 near the Galactic centre massive black hole},
  \emph{\aap}. 615:\penalty0 L15 (July, 2018).
\newblock \doi{10.1051/0004-6361/201833718}.

\bibitem{Neichel+2014}
B.~{Neichel}, J.~R. {Lu}, F.~{Rigaut}, S.~M. {Ammons}, E.~R. {Carrasco}, and
  E.~{Lassalle}, {Astrometric performance of the Gemini multiconjugate adaptive
  optics system in crowded fields}, \emph{\mnras}. {\bf 445}, \penalty0
  500--514 (Nov., 2014).
\newblock \doi{10.1093/mnras/stu1766}.

\bibitem{Unwin08}
S.~C. {Unwin}, M.~{Shao}, A.~M. {Tanner}, R.~J. {Allen}, C.~A. {Beichman},
  D.~{Boboltz}, J.~H. {Catanzarite}, B.~C. {Chaboyer}, D.~R. {Ciardi}, S.~J.
  {Edberg}, A.~L. {Fey}, D.~A. {Fischer}, C.~R. {Gelino}, A.~P. {Gould},
  C.~{Grillmair}, T.~J. {Henry}, K.~V. {Johnston}, K.~J. {Johnston}, D.~L.
  {Jones}, S.~R. {Kulkarni}, N.~M. {Law}, S.~R. {Majewski}, V.~V. {Makarov},
  G.~W. {Marcy}, D.~L. {Meier}, R.~P. {Olling}, X.~{Pan}, R.~J. {Patterson},
  J.~E. {Pitesky}, A.~{Quirrenbach}, S.~B. {Shaklan}, E.~J. {Shaya}, L.~E.
  {Strigari}, J.~A. {Tomsick}, A.~E. {Wehrle}, and G.~{Worthey}, {Taking the
  Measure of the Universe: Precision Astrometry with SIM PlanetQuest},
  \emph{Publications of the Astronomical Society of the Pacific}. {\bf 120},
  \penalty0 38 (Jan., 2008).
\newblock \doi{10.1086/525059}.

\end{thebibliography}
\blankpage
\printindex                         
\end{document}